\documentclass[a4paper,11pt]{article}
\usepackage[cp1251]{inputenc}
\usepackage[dvips]{graphicx}
\usepackage{epsfig}
\begin{document}
\Large

\flushbottom



\textbf{\uppercase{Multicolor Photoelectric $WBVR$ Observations of
the Close  Binary System $\textrm{HZ~Her} = \textrm{Her~X-1}$ in
1986--1988. \protect\hfill\protect\break
I.~Method and Observations}}\\


\textbf{2010  A.N. Sazonov}




{Sternberg Astronomical Institute, Moscow State University,

Universitetskii pr. 13, Moscow, 119992 Russia}


\begin{abstract}
We present results of four-color ($WBVR$) photoelectric observations of the
close binary $\textrm{HZ~Her} = \textrm{HER~X-1}$ in 1986--1988. As a rule,
the duration of the observations exceeded two 35-day X-ray orbital periods in
the 1986--1988 observing seasons. The accuracy and length of the photoelectric
observations facilitated multi-faceted studies, which enabled us to define
several fine photometric effects in the light curves of the binary more
precisely and attempt to interpret them in a model for the matter flow from
the optical component to the accretion disk around the neutron star. This
model provides a satisfactory explanation for the inhomogeneity of the gas flow
and ``hot spot'', as well as the  existence of distinct ``splashes'' moving
in their own Keplerian orbits around the outer parts of the Keplerian disk.
We present series of light curves for all the observing seasons, as well as
color--color diagrams that reflect the physics of various photometric effects.
The transformation coefficients for each of the instrumental systems for the
three observatories at which the observations were carried out are given.
Atmospheric extinction was taken into account during multi-color observations
of the object, with subsequent correction for atmospheric effects with
accuracies ranging from $0.003^m$ to $0.005^m$ for air masses up to $M(z)=2$.
\end{abstract}

\newpage

\section{INTRODUCTION}

The X-ray source Her~X-1 and the physically related optical star HZ~Her
(GSC 2598-01298; $16^{\textrm{h}}57^{\textrm{m}}49.83^{\textrm{s}}$,
${+}35^{\circ}20'32.6''$~$\textrm{{(J2000))}}$, which is an A7 subgiant,
form a close binary system, which exhibits a wide range of unique
and physically distinct optical and X-ray variability~[1].

The X-ray period of the pulsar $P_1 \approx 1.24$~s, which is associated
with the rotation of the neutron star~[2], which is in a disk-accretion
regime, the orbital period  $P_2 \approx 1.7^{\textrm{d}}$ responsible for
the X-ray eclipses and strong optical variability (${\sim} 2^m {-} 3^m)$,
and the precession period $P_3 \cong 34.875^{\textrm{d}}$ due to the
precession of the neutron-star accretion disk, characterize the uniqueness
of the $\textrm{HZ~Her}$ close binary~[3--6]. The X-ray
flux and the shape of the optical light-curve of the close binary vary with
the period $ P_3 \cong 34.875^{\textrm{d}}$~[8, 9].

The optical component periodically eclipses the X-ray source associated with
the neutron star. The main cause of the optical variability of the X-ray binary
is the reflection effect or, more precisely, the heating of the atmosphere of
the optical component (with luminosity $L_v \cong 10^{35}$~erg/s) via
reprocessing of the powerful, hard X-ray emission ($L = 10^{37}$~erg/s at
2--10~keV) in the photosphere of the optical star.

An appreciable asynchronism between the orbital motion and the rotation of the
optical star is observed~[10--12]. Early studies suggested that the optical
star overflows its Roche lobe and loses matter on the thermal time scale.
More recent studies suggest that the dominant effect in the system is
accretion onto the neutron star.

At the same time, this X-ray binary is a classical prototype of an entire
class of X-ray sources with low-mass optical components: the masses of the
X-ray and optical components are $M_x = 1.3 \pm 0.14~M_{\odot}$  and
$M_v = 2.2 \pm 0.1~M_{\odot}$~[13], respectively.

The reflection effect in the close binary does not disappear at any time during
the 35-day precession cycle. This is related to accretion onto the neutron
star and the subsequent formation of an accretion disk in the system~[14].

\begin{table}[htbp]
\caption{Distribution of the observations}

\bigskip
\begin{center}
\begin{tabular}{|l|c|c|c|}
\hline
~~~~~~~~Observatory  &\multicolumn{3}{c|}{Observing season}\\
\cline{2-4}~~~~~~~~~~~(telescope)$^\ast$ &~~~1986  ~~~ &~~~1987 ~~~&~~~1988 ~~~\\
\cline{2-4}&\multicolumn{3}{c|}{Number of nights / Number of individual points in the $WBVR$ bands}\\
&\multicolumn{3}{c|}{}\\
\hline
Crimean Observatory of the SAI       & 14 / 56 & 19 / 149 & 12 / 150  \\
(Zeiss-600 reflector) &&&\\
\hline Maidanak High-Altitude        & 24 / 99 & 10 / 132 & 25 / 183  \\
Observatory of the SAI &&&\\
(Zeiss-600 reflector) &&&\\
\hline Tien Shan High-Altitude       & 43 / 88 &  2 / 24  &  6 / 67   \\
Observatory of the SAI &&&\\
(AZT-14 reflector)&&&\\
\hline
\end{tabular}
\end{center}

\medskip
\hspace{1.2cm} {\footnotesize\mbox{\parbox{10cm}{$^\ast$ An FEY-79
single-channel photometer and standard $WBVR$ filters were used on
all telescopes.}}}

\end{table}

\section{OBSERVATIONS}

 We present here the results of observations of the  close binary
$\textrm{HZ~Her}$ from 1986~[15] to 1988 in the $WBVR$
bands. The $W$ filter ($\lambda_{\textrm{eff}} \approx 3500$~\AA,
$\Delta\lambda_{1 / 2} \approx 520$~\AA) is a revised version of the standard
ultraviolet filter of the $UBVR$ system~[16]. Since the effective wavelength
of the $W$ filter is ${\sim} 100$~\AA{} shorter than the wavelength of the
$U$ filter, color variations in the close-binary system are expressed more
strongly in our observations than in $UBV$ data. The total number of individual
$WBVR$ observations carried out at three observatories over 155 nights in
1986--1988  is 948 (Table~1).

 We used $\textrm{C3} = \textrm{GSC~2598-01270}$ ($16^{\textrm{h}}57^{\textrm{m}}
17.84^{\textrm{s}}$, ${+}35^{\circ}21'45.0''$,  J2000) as a comparison star
in these observations. We obtained the $WBVR$ magnitudes and color indices of
this star by matching to the standards HD~152380, HD~147924, HD~148253~[14]:

$$
\begin{array}{cc}
W = 12.920^m \pm 0.050^m,&\quad B = 13.172^m \pm 0.032^m,\\
V = 12.596^m \pm 0.014^m,&\quad R = 12.127^m \pm 0.020^m.
\end{array}
$$


 These values agree with the $B$ and $V$ magnitudes of C3 found
in~[17] within the errors.

 The stars $\textrm{C2} = \textrm{GSC~2598-01267}$ ($16^{\textrm{h}}
57^{\textrm{m}}34.41^{\textrm{s}}$, ${+}35^{\circ}21'57.0''$,  J2000) and
$\textrm{C4} = \textrm{GSC~2598-01274}$ ($16^{\textrm{h}}58^{\textrm{m}}
06.21^{\textrm{s}}$, ${+}35^{\circ}21'32.9''$,  J2000) were also used as
a comparison star and check star, respectively. To reduce the errors of the
photoelectric observations, a comparison star was observed in the same filter
before and after every observation of $\textrm{HZ~Her}$. The check star was
observed two to four times during each observing session. The background in
the vicinity of $\textrm{HZ~Her}$ was also periodically measured.

 To obtain a denser set of observations, the comparison star C3 and background
were measured every 30--40~min with subsequent extrapolation to the time of
the observations of $\textrm{HZ~Her}$.  As in~[18], the transformation
coefficients were also measured each season. Because the instrumental system
we used differs only slightly from the standard  $UBVR$ system, the relation
between the transformation coefficients can be expressed by the first-order
linear equations

$$
\begin{array}{c}
V=v_0+\eta_v+\xi_v\, (B-V),\\
U-B= \eta_{U-B}+\xi_{U-B}\,(u-b)_0,\\
B-V= \eta_{B-V}+\xi_{B-V}\, (b-v)_0,\\
V-R=\eta_{V-R}+\xi_{V-R}\, (v-r)_0,\\
\end{array}
$$


 where the unknowns are the transformation coefficients $\eta$ and
$\xi$. The best photometric nights were used to determine these
coefficients. The transformation coefficients $\xi$ were then
averaged during each observing season, and the values of the
zero-points $\eta$ for each night of the given season were
computed using these average $\xi$.

\begin{table}[htbp]
\caption{Transformation coefficients 
$\xi$}

\bigskip

\hspace{0.45cm}
\begin{tabular}{|l|c|c|c|c|c|} \hline
~~~~~~~~Observatory &$\xi_{V}$&$\xi_{W-B}$&$\xi_{B-V}$&$\xi_{V-R}$&$n$\\
 \hline
Crimean Observatory of the SAI &$0.013\pm 0.003$& $0.962\pm 0.005$& $1.102\pm 0.003$& $1.088\pm 0.004$& 38\\
\hline Maidanak High-Altitude  &$0.012\pm 0.003$& $0.958\pm 0.004$& $0.937\pm 0.007$& $1.065\pm 0.007$& 41\\
Observatory SAI &&&&&\\
\hline Tien Shan High-Altitude &$0.054\pm 0.002$& $0.997\pm 0.009$& $0.929\pm 0.005$& $1.068\pm 0.008$& 27\\
Observatory SAI &&&&&\\
\hline
\end{tabular}
\end{table}




 The calculated average values of $\xi$ and their errors are listed in Table~2.
In Table~2, $n$ is the number of nights used to derive $\xi$. We always used
the same detectors on the Zeiss-600 and AZT-14 reflectors. Our detector was
a FEU-79 (multi-alkali photo-cathode S-20) photomultiplier. If the
photomultiplier was changed, the transformation coefficients $\eta$ and $\xi$
were recalculated.

 Atmospheric extinction was taken into account during broad-band (multi-color)
observations of HZ~Her (including the $WBVR$ observations). Moreover, the
color characteristics of HZ~Her change substantially during the orbital and
precession periods. During differential observations of variable stars using
standard and check stars that were similar in color and located close in the
sky, the problem of atmospheric extinction essentially does not arise. Account
for atmospheric extinction is also necessary to reduce the systematic
measurement errors to values of the order of $0.001^m{-}0.002^m$.

 Assigning a standard spectral energy distribution is still a problem for
peculiar stars such as HZ~Her; in this case, iteratively correcting for
atmospheric extinction in fundamental multi-color astrophotometry, as is
proposed in~[19], enables correction for atmospheric effects for our broad-band
$WBVR$ measurements with accuracies no worse than  $0.005^m$ in $W$ and
$0.003^m$ in $B$, $V$, and $R$, if the air mass $M(z)$ does not exceed two
at the high-altitude observatories ($h\geq3000$~m). These errors must be
approximately doubled for the lower-altitude observatories~[19]. We are speaking
here of systematic errors. The random errors (atmospheric flickering, rapid
variations of the transparency, photon noise in the detectors etc.) can
be larger, but they can be suppressed by increasing the number of independent
measurements.

 During observations made together with T.R.~Isrambetova at the
Maidanak High-Altitude Observatory (Uzbekistan) in 1987--1988, we
used the reduction coefficients of the Zeiss-600 telescope (a
single-channel $WBVR$ electrophotometer with an automatized
control system, using a FEY-79 photomultiplier as a detector). The
relative spectral sensitivity of the system was fairly stable, and
was checked twice per season (Spring-Summer and Summer--Autumn).
The reduction coefficients to the standard photometric system were
obtained using multiple measurements of the standard stars in the
fields SA~107, 108, and 111-113~[20]. This yielded the relations

$$
\begin{array}{c}
B{-}V=1.071(\pm0.021)(b{-}v)-0.068^m(\pm0.018^m),\\
V{-}R=0.803(\pm0.033)(v{-}r)+0.173^m(\pm0.014^m),
\end{array}
$$

 where $b$, $v$, and $r$ are the instrumental magnitudes and $B$,
$V$, and $R$ are the magnitudes in the Johnson photometric system.

 We calculated the corrections to the magnitudes using the known instrumental
color indices using the expressions

$$
\begin{array}{c}
B{-}b=0.094(b{-}v)-0.099^m,\\
V{-}v=0.014(b{-}v)-0.015^m,\\
R{-}r=0.234(v{-}r)-0.208^m.
\end{array}
$$


\begin{table}[htbp]
\caption{Results of the observations}

\begin{center}


\begin{tabular} {|p{20pt}|p{70pt}|p{35pt}|p{35pt}|p{35pt}|p{35pt}|p{35pt}|p{35pt}|p{20pt}|}
\hline  &JD 2400000+ & ~~~~$\varphi$&~~~~$\psi$ &
~~~~$W$& ~~~~$B$ & ~~~~$V$ & ~~~~$R$ &  ~~$n$ \\
\hline
~~~~1&~~~~~~~~2&~~~~3&~~~~4&~~~~5&~~~~6&~~~~7&~~~~8&~~9\\\hline 1&
46615.3830 & 0.873 & 0.756 &15.075 & 15.125
& 14.667 & 14.328 & ~~6 \\
\hline 2& 46616.4574 & 0.983 & 0.757 & 15.144 &  15.314
& 14.946 & 14.660 & ~~8 \\
\hline 3& 46618.4099 & 0.181 & 0.758 & 14.855 &  15.306
& 14.779 & 14.585 & ~~6 \\
\hline
\end{tabular}
\end{center}
\label{tab1}
\end{table}




 In the later observations in 1989--1998 (and in all observatories at which
observations were carried out by the author), the reduction coefficients to
the standard Johnson photometric system were obtained from measurements of
standard stars in the field of h and $\chi$~Per (12 standard stars were
measured each time) and of NGC~884 (13 standard stars were measured each time).

 The scatter of the individual points on the light curves is substantially
larger than the observational errors. This probably reflects physical
variability of the system~[21, 22].

 The observations were processed using a differential method. The orbital
phases $\varphi$ were calculated according to the elements~[23]

$$
\begin{array}{c}
\textrm{Min Hel} = \textrm{JD} 2441329.57519  +
1.70016773^{\textrm{d}} {}E.
\end{array}
$$

 The observational data are summarized in Table~3. We present part of the table
as an illustration; the entire table is located at
(http://lnfm1.sai.msu.ru/$\sim$sazonov/). The columns of Table~3 give (1)
the consecutive number of each observation, (2) the Julian dates of the
observations, (3) the phases $\varphi $ of the orbital period, (4) the
precession phases  $\psi_{35}$, (5)--(8) the magnitudes in the $W$,  $B$,
$V$, and $R$ bands, and (9) the number of individual observations  $n$.

 The observational data for HZ~Her for all years were analyzed according to the
phase of the 35-day cycle, taking into account the fact that precession times
correspond to the ephemeris~[24]

$$
\begin{array}{c}
  T_{35}={\textrm{JD}} (2441781.0{\pm 0.5)} + (34.875^{\textrm{d}}{\pm0.003^{\textrm{d}}})  E.
\end{array}
$$

 The phases $\psi _{35}$ of the 35-day cycle were calculated
assuming that $\psi _{35}=0$ when the X-ray source turns on~[14,
24--27].

\begin{figure*}[p!]
\includegraphics{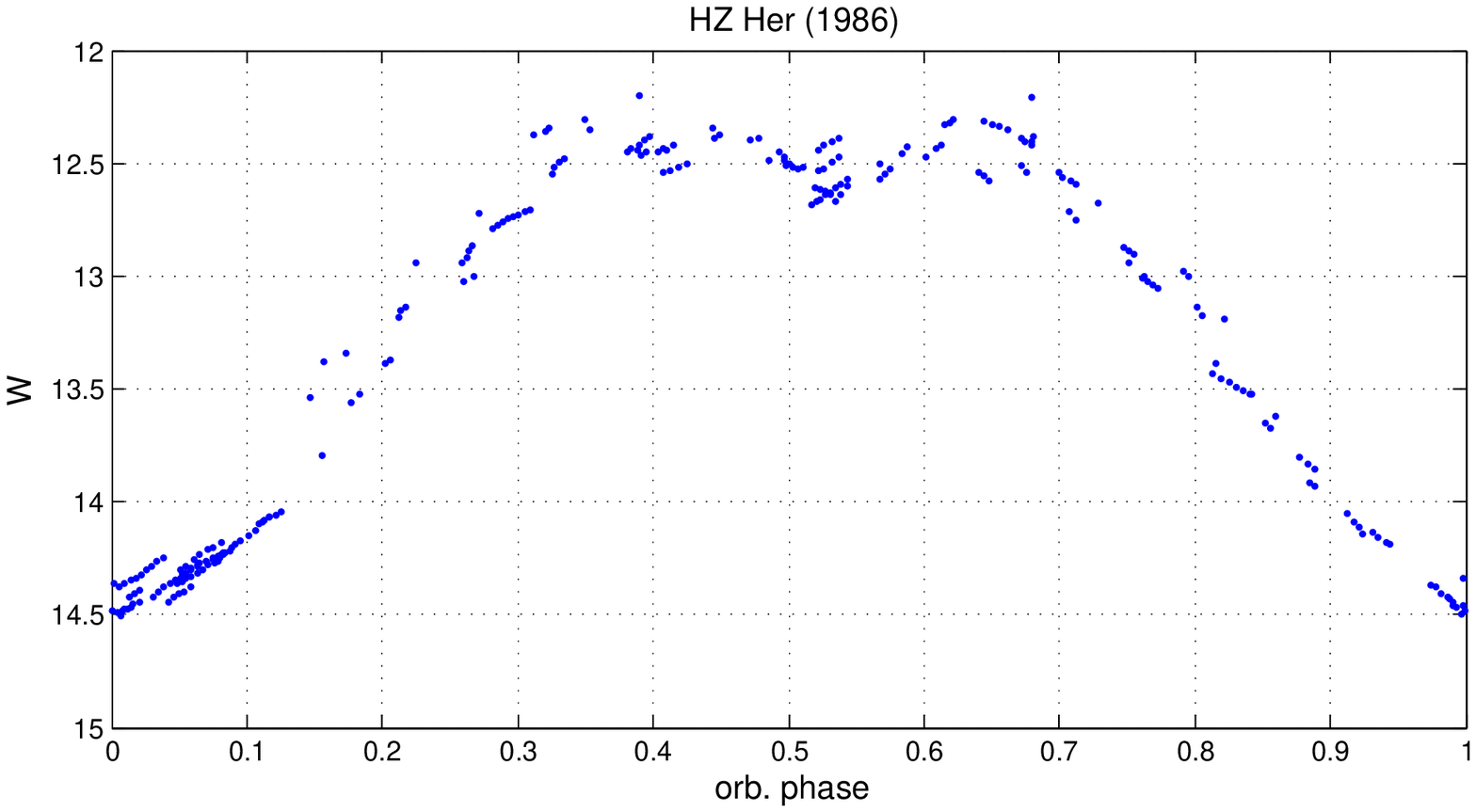}

\caption{Light curve of  HZ~Her in 1986 in  (a) $W$, (b)  $B$, (c) $V$, and
(d) $R$ convolved with the orbital period $P=1.70016773^{\textrm{d}}$.
\hfill}
\end{figure*}

\section{OBSERVATIONS IN 1986}

 $WBVR$ observations of the $\textrm{HZ~Her} = \textrm{Her~X-1}$
close binary were carried out in July--October 1986 using the
600-mm reflector of the Crimean Station of the Sternberg
Astronomical Institute (SAI), the 480-mm (AZT-14) reflector of the
Tien Shan High-Altitude Observatory, and the 600-mm reflector of
the Maidanak High-Altitude observatory, using a single-channel
electrophotometer in a photon-counting regime (Fig.~(1.1-1.4)). In
1986, the reflection effect in the system remained at the
classical (average) level, as can be seen from the light curves
(Figs.~1-4; this and all other figures can be found in electronic
form at http://lnfm1.sai.msu.ru/$\sim$sazonov/HZ~Her$=$Her~X-1).
The main origin of this optical variability of HZ~Her is the
reprocessing of X-rays in the photosphere of the optical component
of the system~[11].

 The spatial position of the X-ray source Her~X-1 relative to the optical
component HZ~Her is always known with high precision. Therefore, we always
have comprehensive information about the shape of the hot spot in the
photosphere of the optical star, and, hence, about the spatial distribution of
the X-ray emission of the neutron star.

\begin{figure*}[p!]
\includegraphics{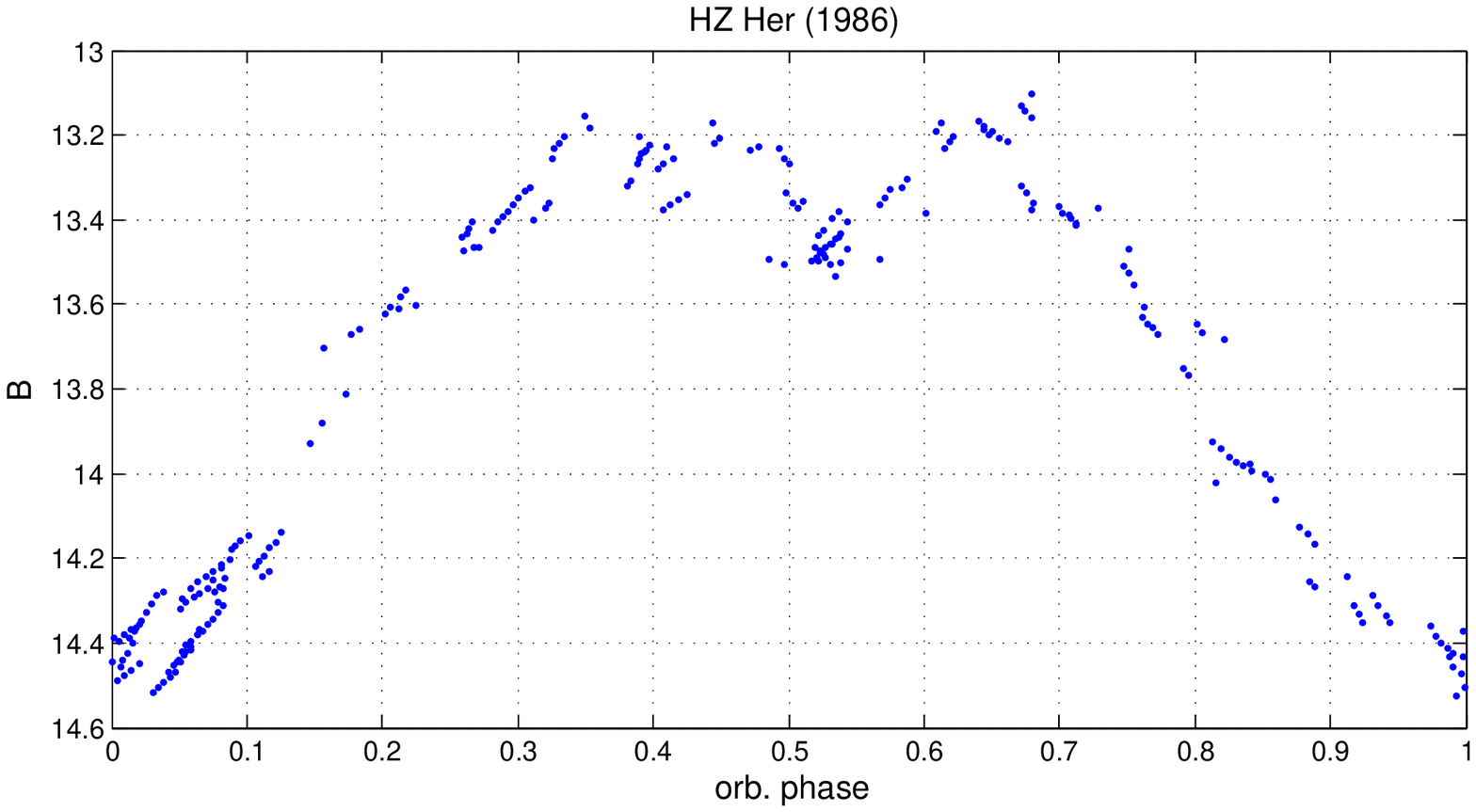}

\caption{Short flares of emission close to orbital phases
0.015--0.025 in the (a) $W$, (b) $B$,  (c) $V$, and (d) $R$ bands.
\hfill}
\end{figure*}

\begin{figure*}[p!]
\includegraphics{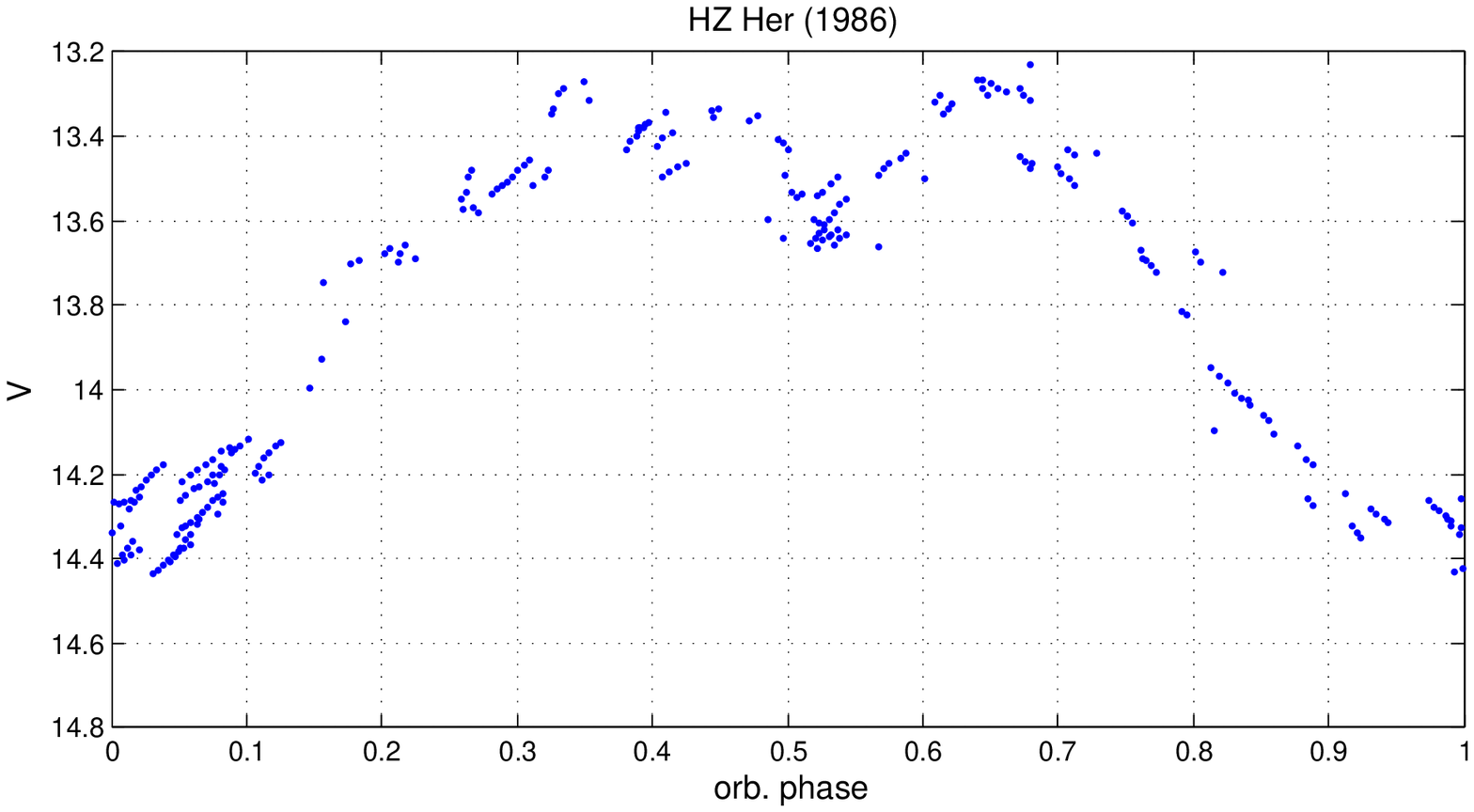}

\caption{Diagrams of (a) $(W{-}B)$, (b) $(B{-}V)$ , (c) $(V{-}R)$,
and (d) $(B{-}R)$ vs. orbital phase $\varphi$ for the 1986 season.
\hfill}
\end{figure*}

\begin{figure*}[p!]
\includegraphics{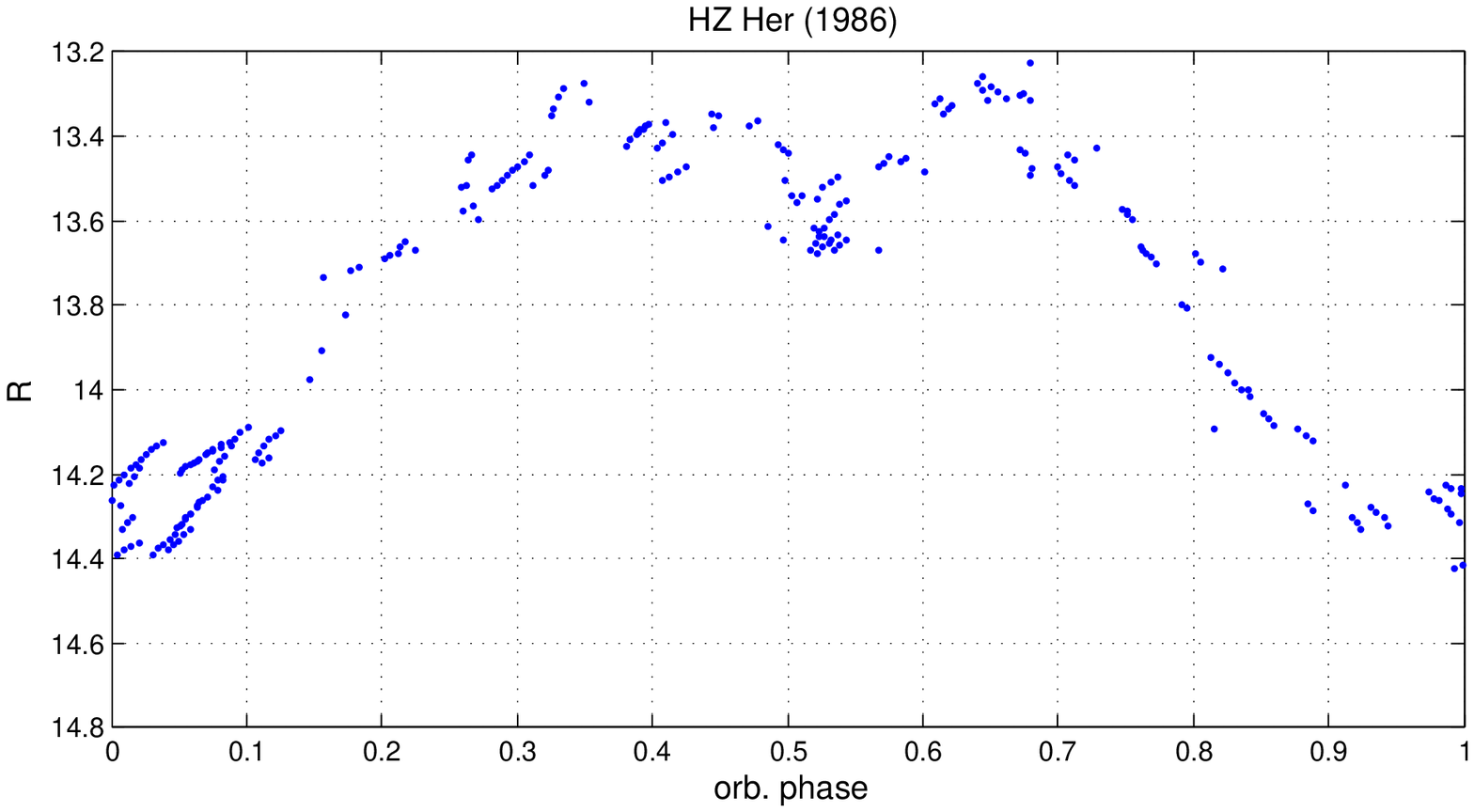}

\caption{Diagrams of (a) $(W{-}B)$, (b) $(B{-}V)$ , (c) $(V{-}R)$,
and (d) $(B{-}R)$ vs. orbital phase $\varphi$ for the 1986 season.
\hfill}
\end{figure*}

 In  models with an anisotropic X-ray source with a complex accretion disk
around the neutron star~[28, 29], a certain asymmetry of the hot spot forms,
which is reflected as ``shoulders'' with different heights in the  optical
light curves (in our chosen theoretical formalism, ``shoulders'' are  those
parts of the light curve at orbital phases from $\varphi=0.40$ to
$\varphi=0.57$).

 In the 1986 season, short flares (10 to 20 minutes long) were
detected during individual nights in $W$,  $B$, $V$, and  $R$
close to orbital phases 0.015--0.025, with amplitudes of $0.03^m$,
$0.02^m$, $0.02^m$, and $0.01^m$, respectively (Fig.~(2.1-2.4).

 We are interested in analyzing the observed manifestations of
accretion-related structures in the system (most importantly, the
accretion disk around the neutron star, gas condensations and
other perturbing components) relative to the minimum brightness in
the $W$, $B$, $V$, $R$ bands (Fig.~(1.1-1.4) and the minimum
$W{-}B$ color index (Fig.~(3.1-3.4), close to the main minimum
(orbital phases $\varphi=0.96{-}0.04$). The brightness and $W{-}B$
variations (Fig.~3.1) are most probably due to precession and the
physical parameters of the neutron-star accretion disk.

 Characteristic kinks in light curves are observed at these same orbital
phases~[30]. There is a temporal correlation of these kinks in the optical
light curve with the formation of dips in the X-ray curve. The latter are
probably related to a certain regime for the mass flow from the optical
component of HZ~Her onto the neutron star~[27]. These typical kinks and
inflection points in the vicinity of the primary minimum (Min~I) are essentially
present in the 1986 light curves at all precession phases of the 35-day cycle.

 All these effects lead to complex evolution of the optical light curves between
observing seasons. The light curves in all spectral bands have a sharp bell-like
shape, with a wide Min~I and well-defined secondary minimum (Min~II), but
somewhat higher-amplitude right branch of the light curve (close to orbital
phase $\varphi = 0.63$). This photometric peculiarity provides evidence for a
displacement of the hot spot with orbital phase (with its area constant in the
1986 season), relative to the central meridian of the optical component
of the system.

\begin{figure*}[p!]
\includegraphics{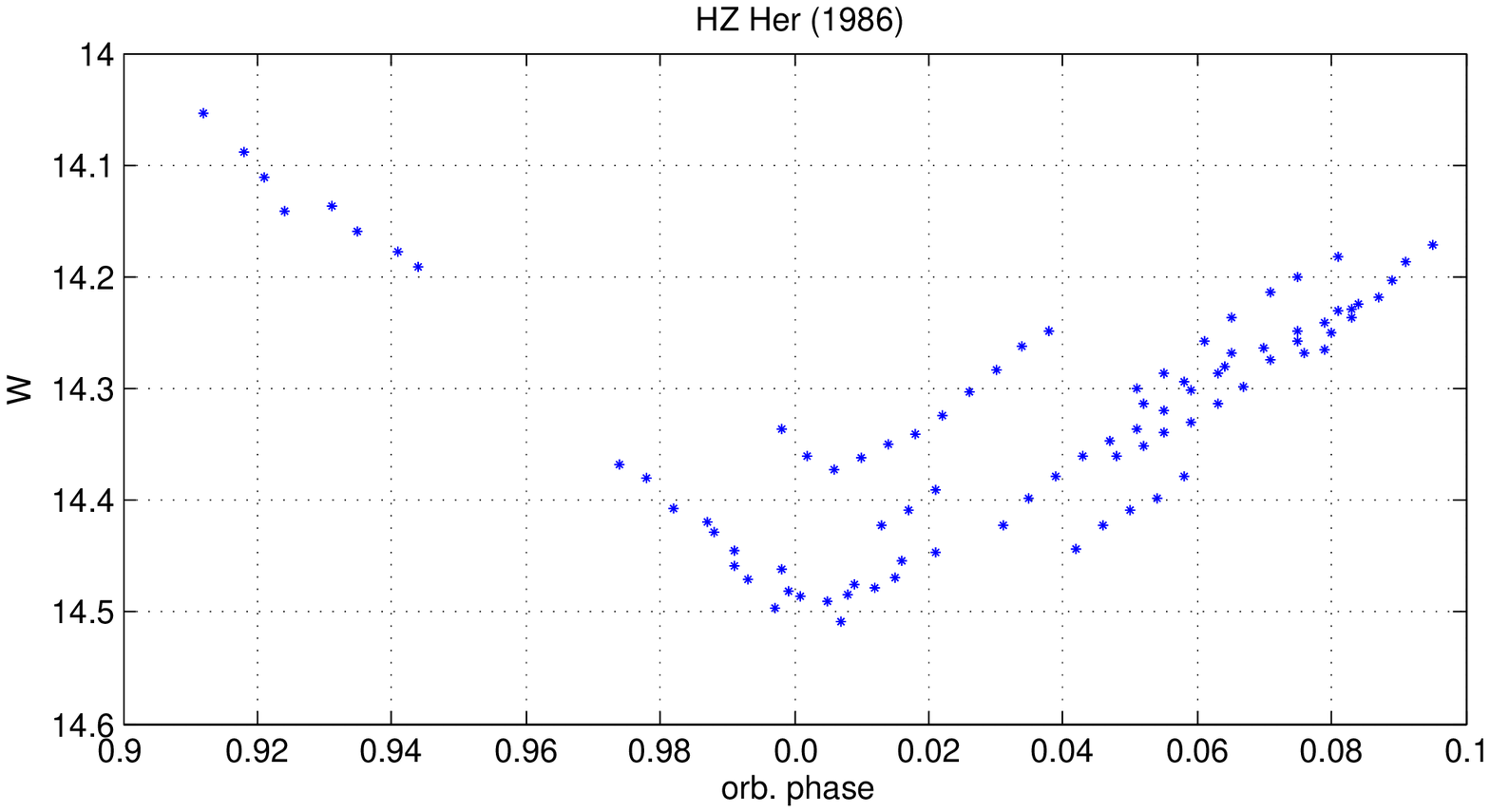}

\caption{Light curve of  HZ~Her in 1986 in  $W$ for Min I.
\hfill}
\end{figure*}

\begin{figure*}[p!]
\includegraphics{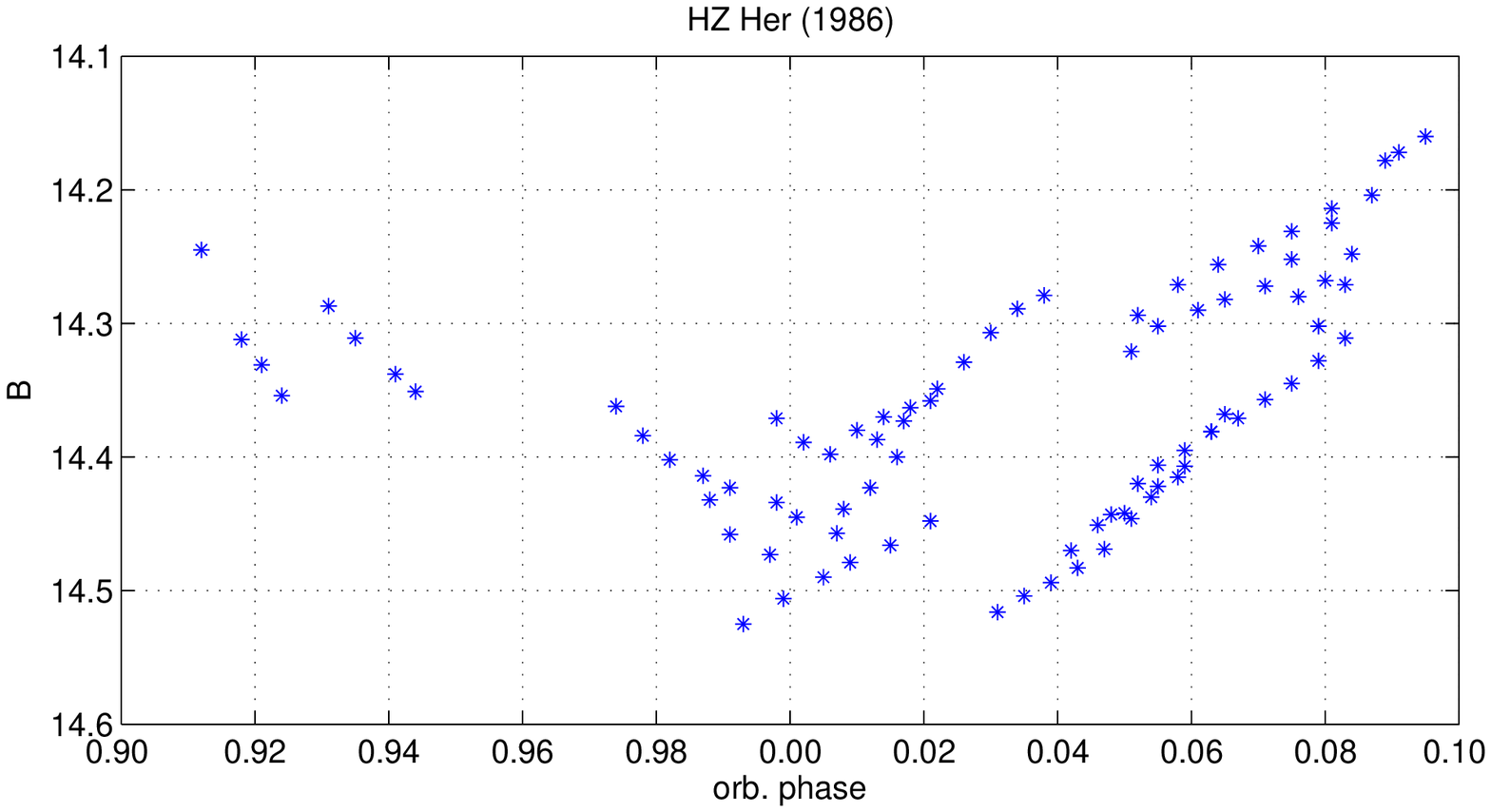}

\caption{Light curve of  HZ~Her in 1986 in  $B$ for Min I.
\hfill}
\end{figure*}

\begin{figure*}[p!]
\includegraphics{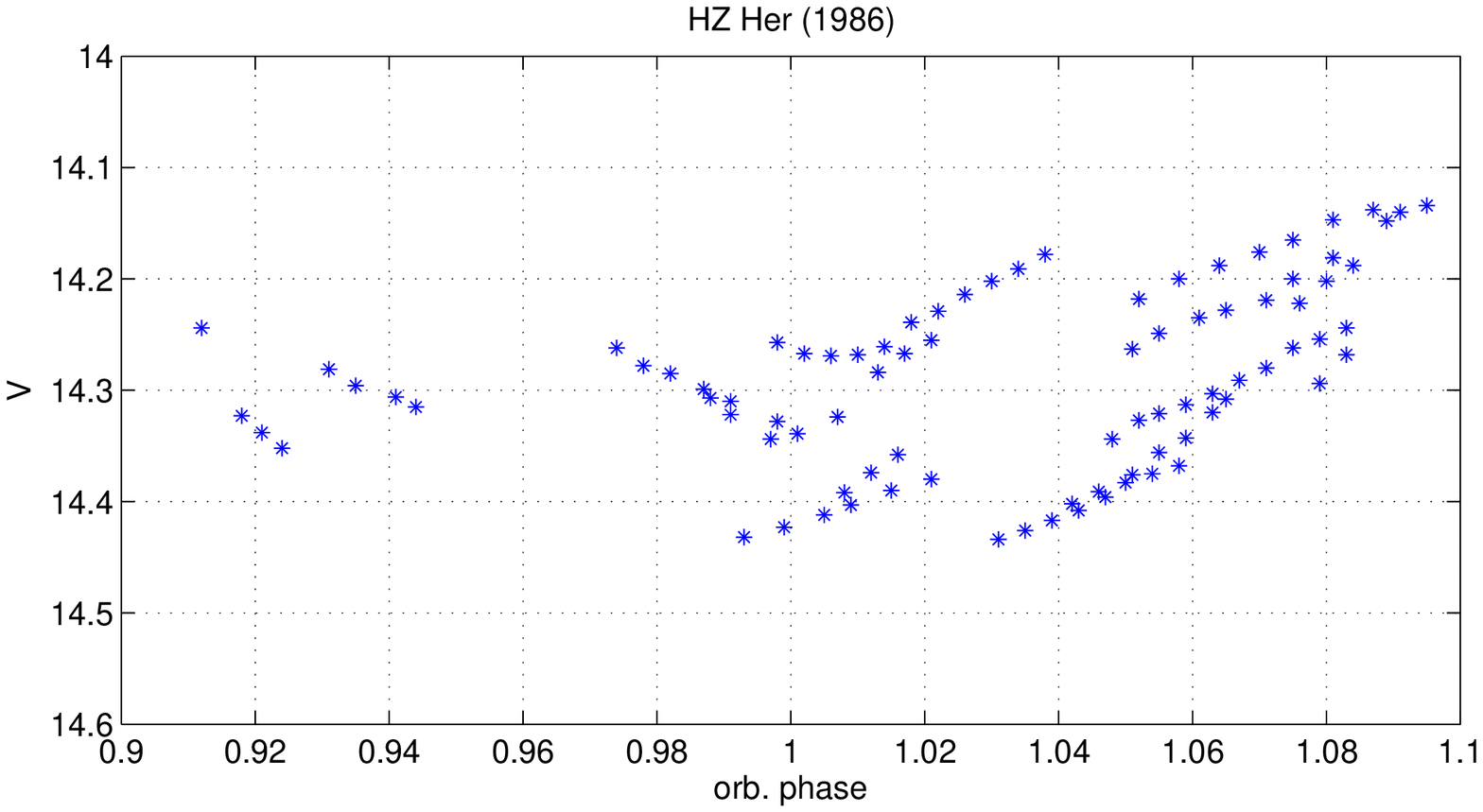}

\caption{Light curve of  HZ~Her in 1986 in  $V$ for Min I. \hfill}
\end{figure*}

\begin{figure*}[p!]
\includegraphics{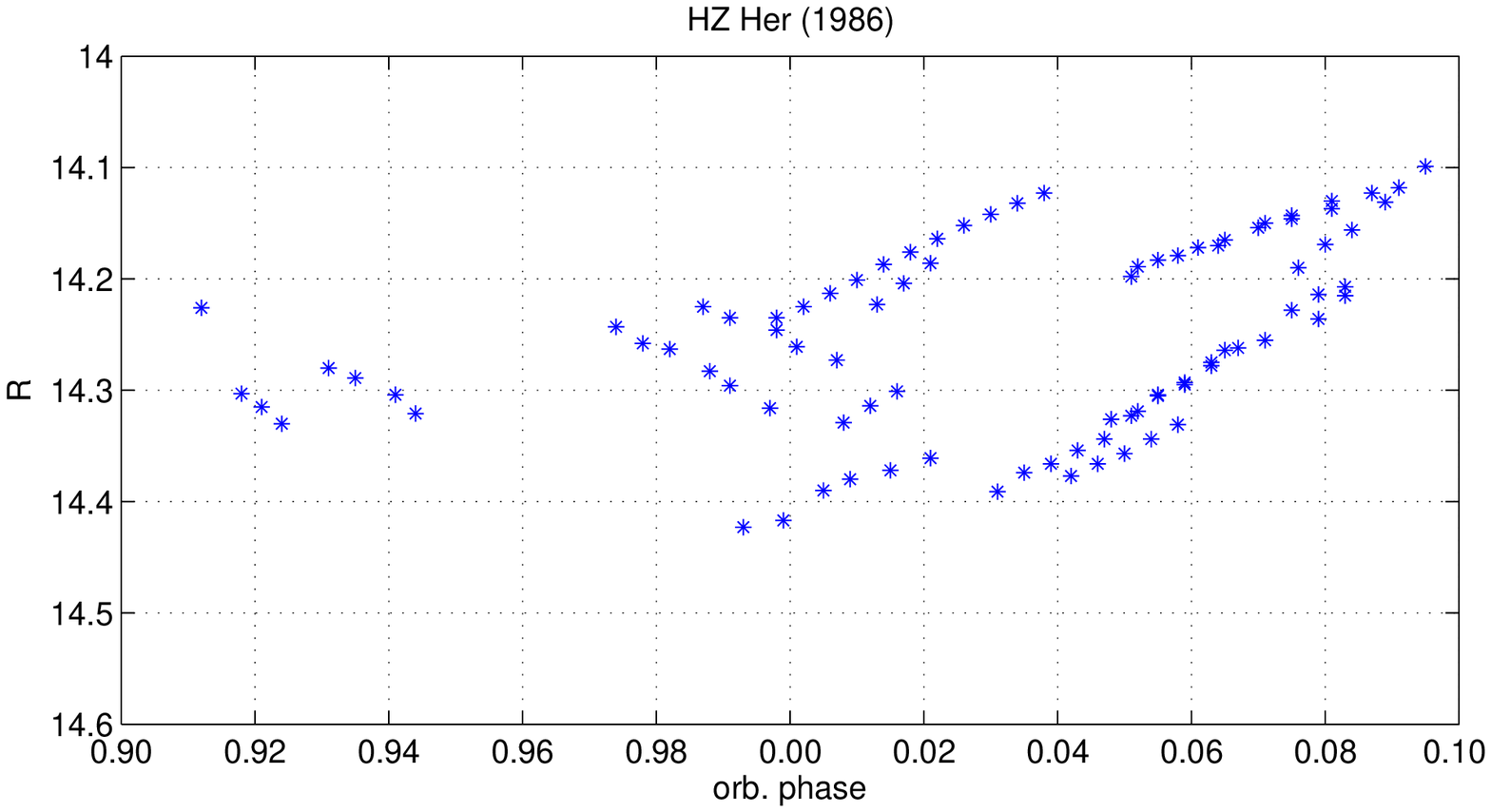}
\caption{Light curve of  HZ~Her in 1986 in  $R$ for Min I.
\hfill}
\end{figure*}

 At the Min~I ingress in the 1986 season, the light curve is
steeper and more ordered and ``classical''{} during the decline,
and flatter during the rise, displaying a small dispersion of the
brightnesses at the same orbital phases measured on different
nights. This observational data indicate some asymmetry of the
light curve close to the primary minimum. This suggests a
variation of the colorimetric parameters of the accretion disk
when it is eclipsed by the optical star. The difference in the
brightness of the accretion disk at the ingress and egress of the
X-ray eclipse is $0.20^m$ (Fig.~1.1), consistent with results of
similar computations made in~[31].

 The color indices of the outer regions of the accretion disk
differ quite significantly at the ingress and egress of the X-ray
eclipse (Fig.~(3.1-3.4)). The same is true of the
$(W{-}B)-(B{-}V)$ and $(V{-}R)-(B{-}V)$ color--color diagrams
(Fig.~(4.1-4.2).

 Based on these observational data, we conclude that there was an increase in
the size of the accretion structures in HZ~Her in 1986~[27, 31], which is
reflected in the shape of the light curve (the right shoulder of the light
curve is somewhat higher than the left one).

 Moreover, close to Min~I (orbital phases $\varphi=0.96{-}0.04$), the variations
of the minimum brightness and $W{-}B$ were of the order of $0.25^m$, which
happens in this system extremely rarely~[31--33].

 These variations in the brightness (Fig.~(1.1-1.4)) and color index (Fig.~3) with phase
$\varphi $ are probably most plausibly explained by variations in the size of
the accretion structures in the system, which re-radiate the X-ray flux in
the optical~[14, 31, 34].

 In addition, the system probably loses mass through the second Lagrange point
{L$_2$}. All this facilitates the formation of extended accretion structures in
the system and their manifestation in the optical at all phases of the 35-day
precessional cycle.

 Also in 1986, significant changes in the physical conditions in the
$\textrm{HZ~Her = Her~X-1}$ close binary caused some observed effects that were
correlated in the X-ray and optical.

 According to EXOSAT data~[35--37], nonuniform X-ray irradiation of the outer
regions of the accretion disk is observed at certain phases of the precessional
cycle, and asymmetry of the reprocessed disk X-ray emission is observed in the
optical. This corresponds to the model proposed in~[38, 39] and elaborated
in~[31, 32].

 Thus, there are reasons to suggest that there is an increase in the accretion
rate onto the neutron star in this case. The first reason for this could be
a strong increase (decrease) in the mass influx into the neutron-star accretion
disk, related to different mass-flow regimes and conditions for the outflow from
the Lagrange point L$_1$ at different epochs~[32, 40]. Second, the viscosity of
the accretion disk could suddenly increase, causing it to become turbulent
rather than laminar~[41], or this may be a manifestation of a turbulent
accumulating disk~[42]. All these effects are reflected in the 1986 light curve.
Third, note also that, at the orbital phases of elongations of the X-ray source
($\varphi=0.25$ and $\varphi=0.75$), the light curve of the system is
practically the same as light curves obtained at other epochs in the same
season; this indicates that the area of the hot spot was approximately constant
during the 1986 season.

 The main principles we have used in our analysis of the light curve to
determine the geometry and spatial location of the hot spot, as well as the
physical parameters of the accretion structures in the system, are presented
in~[39]. When applied to the observational data, these mathematical methods
enable us to fit the data and to find the reflection of various effects in the
light curves obtained in 1986.

 Fitting of the data obtained close to the X-ray eclipses suggests that the
brightness of the system was somewhat higher than in 1987--1988 and 1989--1998
(for which there are still unpublished photoelectric observations).

 The $W{-}B$ color index was also higher in 1986 than in the following years.
Analysis of the $(W{-}B) {-} (B{-}V)$, $(V{-}R) {-} (B{-}V)$
color--color diagrams (Fig.~3.1, 3.2) leads to the logical
conclusion that the hot spot was located on the trajectory of the
motion of the accretion structures along the limb of the optical
component of the close binary.

\begin{figure*}[p!]
\includegraphics{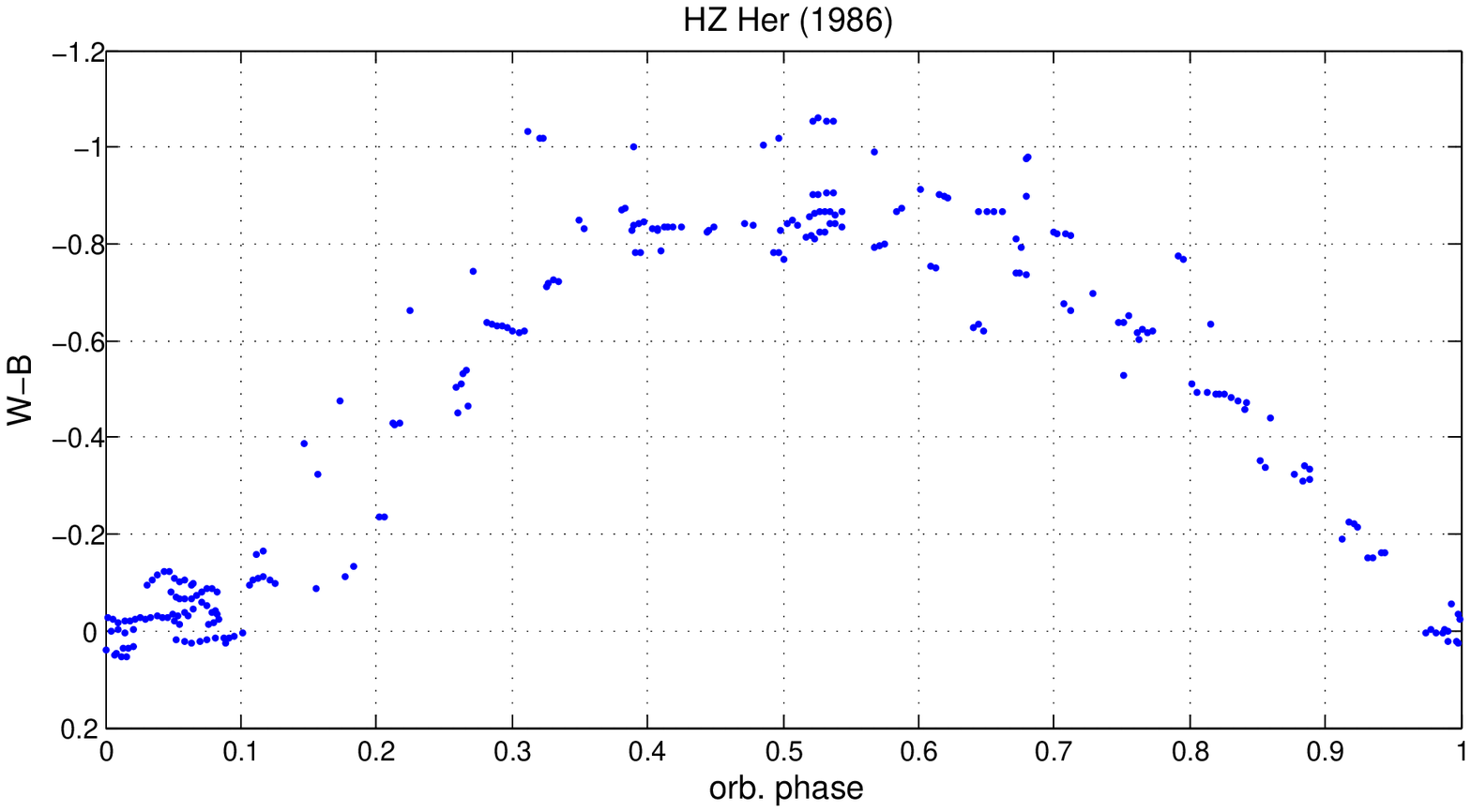}
\caption{Same as Fig.~1 for 1986.
\hfill}
\end{figure*}

\begin{figure*}[p!]
\includegraphics{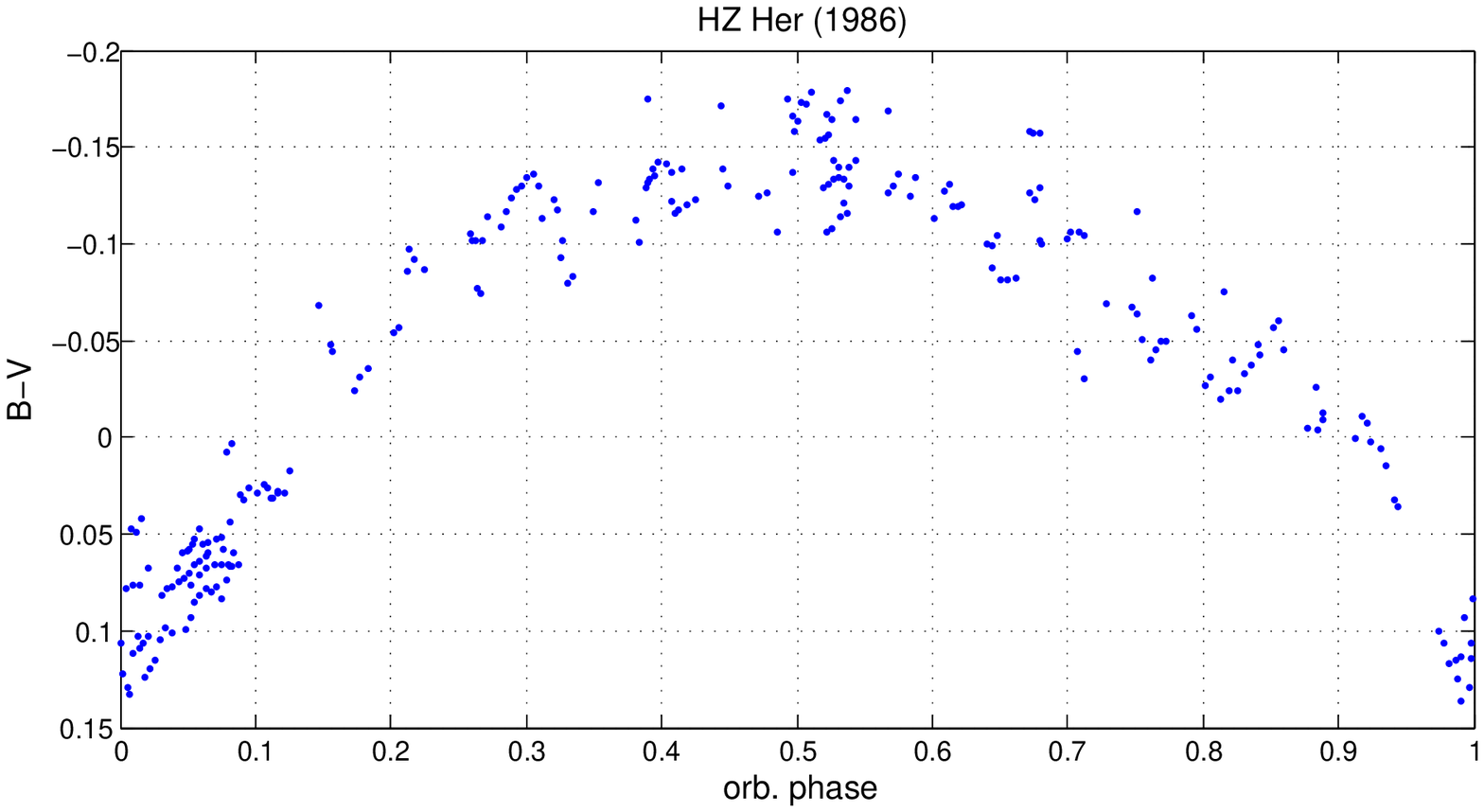}
\caption{Same as Fig.~1 for 1986. \hfill}
\end{figure*}

\begin{figure*}[p!]
\includegraphics{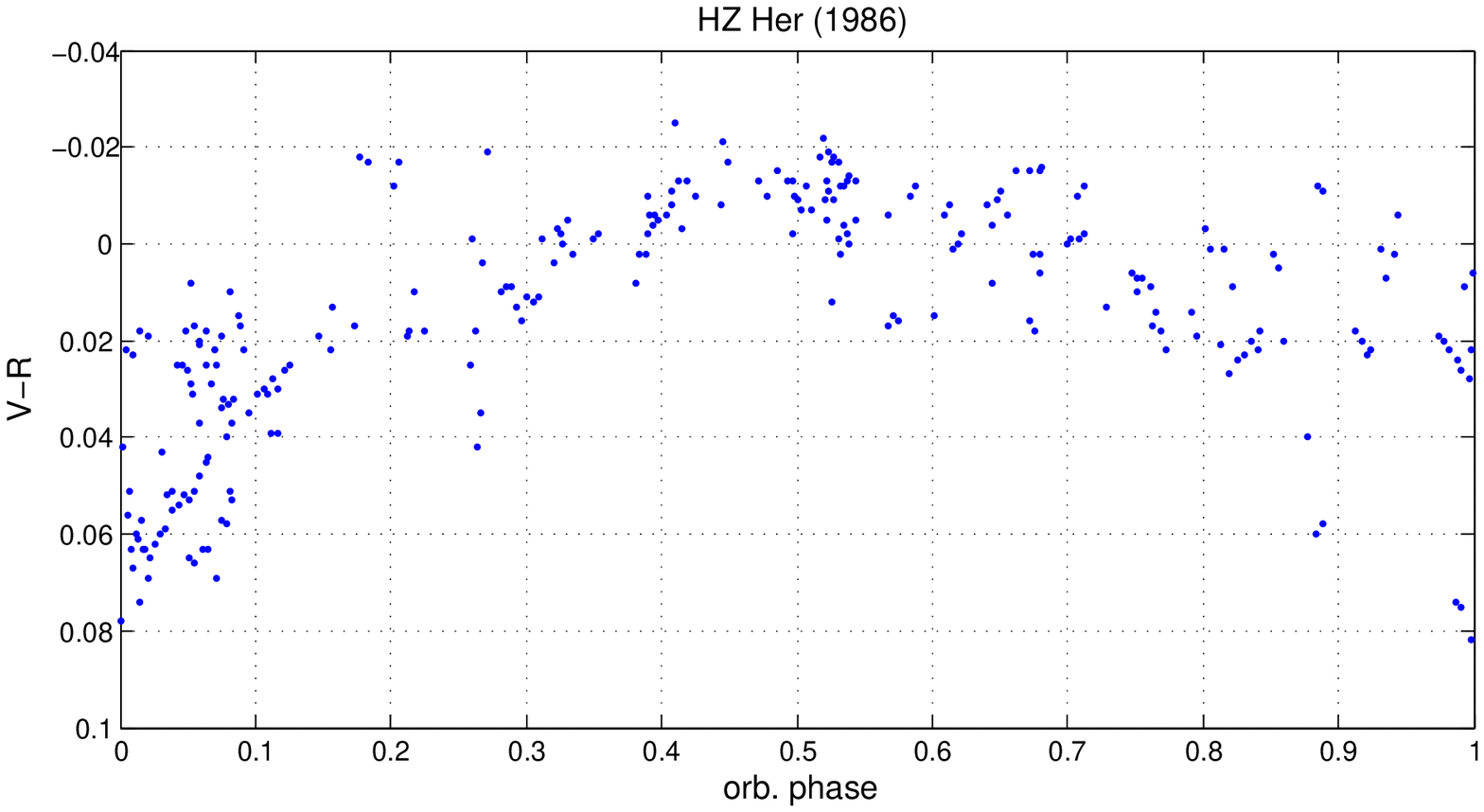}
\caption{Same as Fig.~1 for 1986. \hfill}
\end{figure*}

\begin{figure*}[p!]
\includegraphics{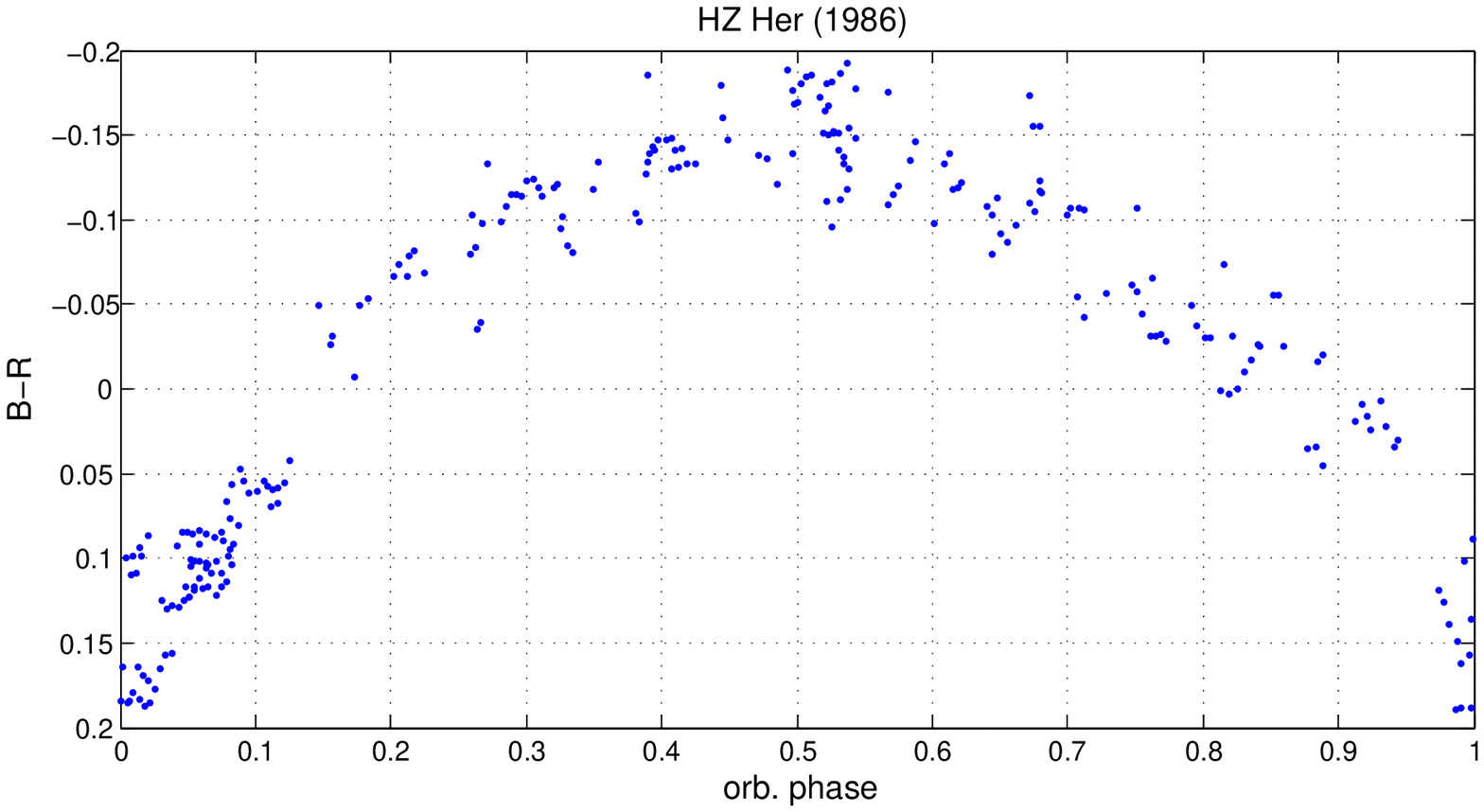}
\caption{Same as Fig.~1 for 1986. \hfill}
\end{figure*}

\begin{figure*}[p!]
\includegraphics{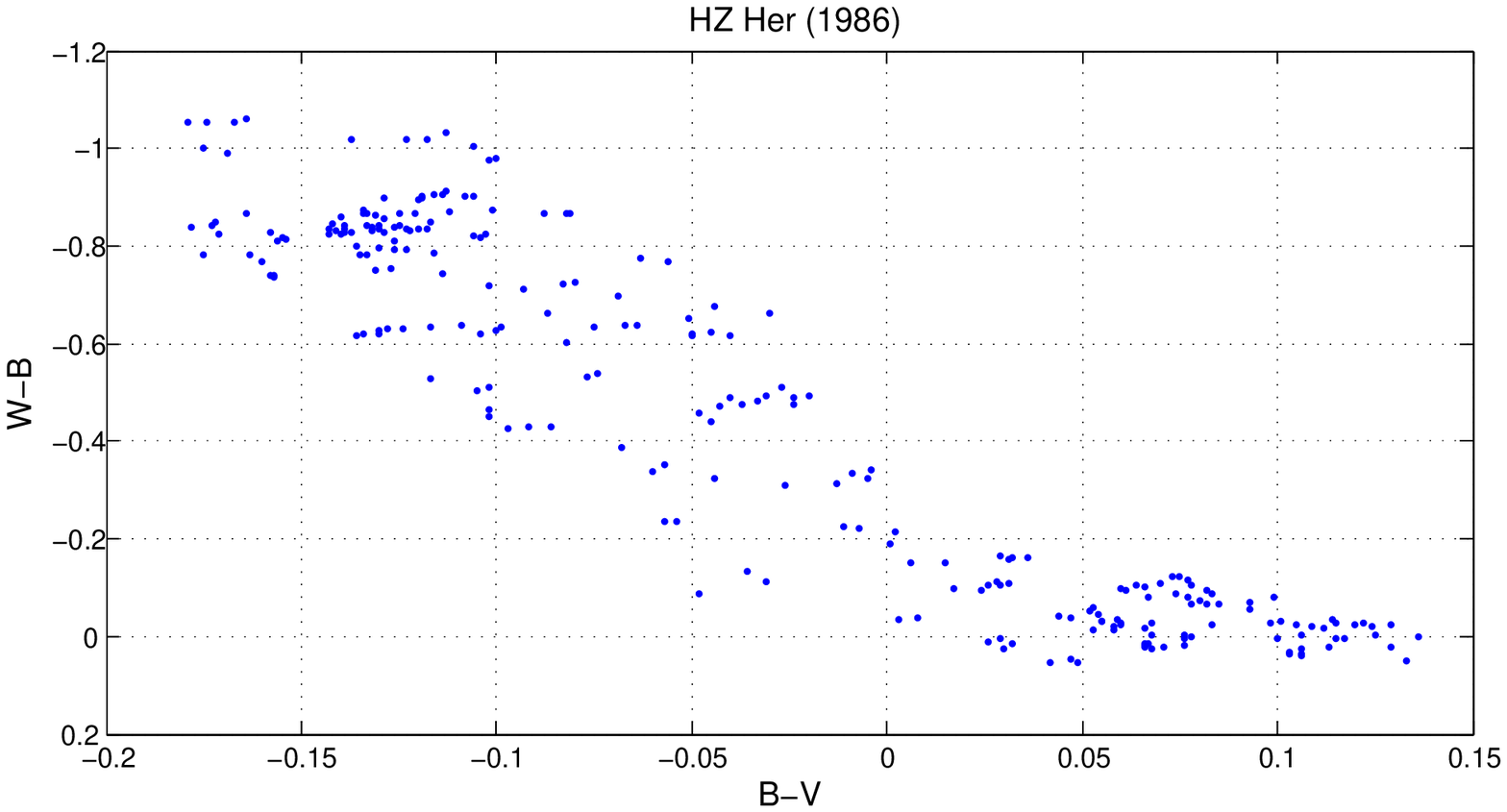}
\caption{Color--color diagrams (a) $(W{-}B)-(B{-}V)$ for the
1986~season.
\hfill}
\end{figure*}

\begin{figure*}[p!]
\includegraphics{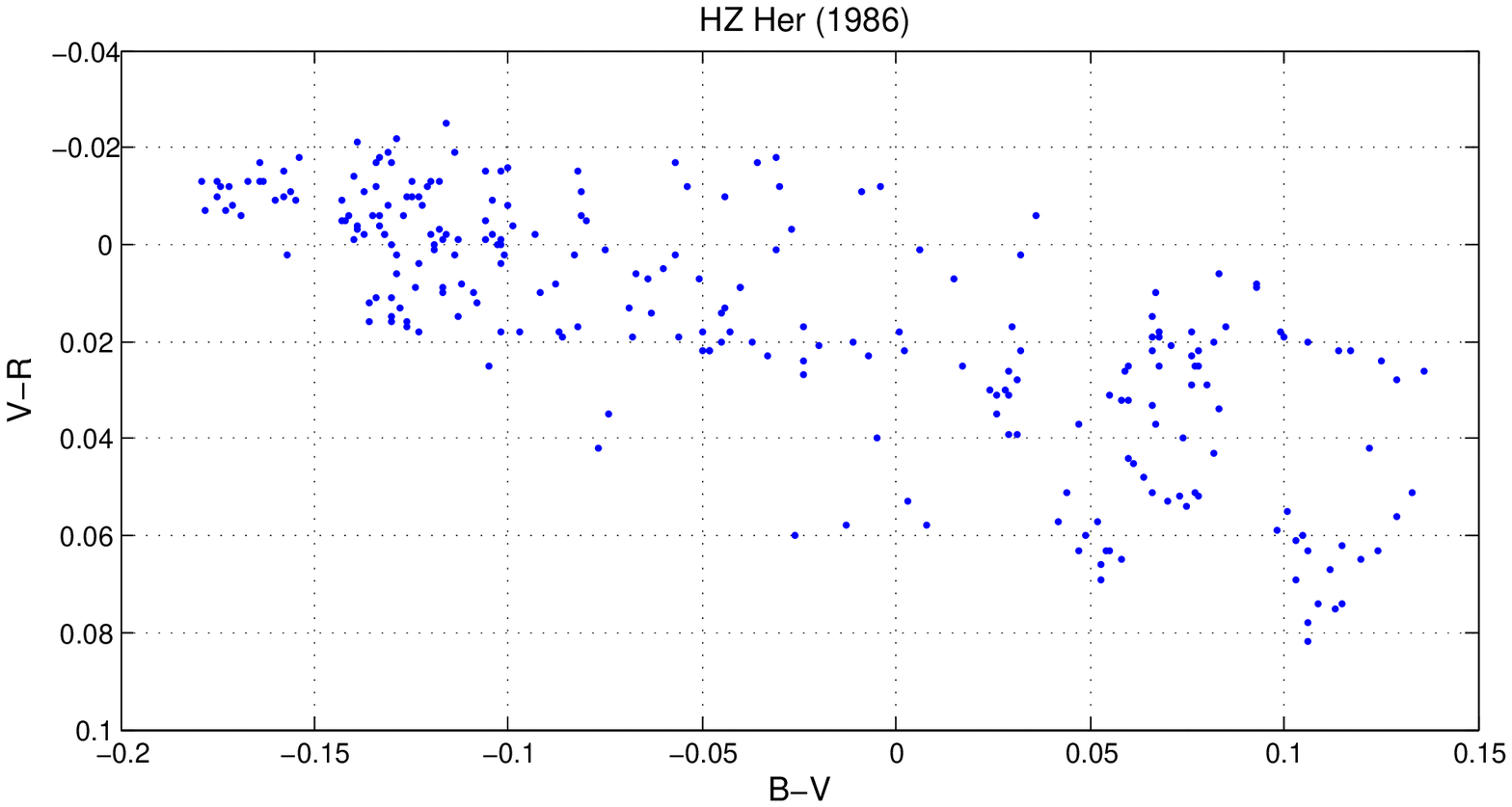}
\caption{Color--color diagrams $(V{-}R)-(B{-}V)$ for the
1986~season.
\hfill}
\end{figure*}

\section{OBSERVATIONS IN 1987}

 The results of the 1987 observations are presented in Fig.~(5.1-5.4). As is more
common, the left shoulder of the secondary minimum on the light
curve is higher than the right one. Moreover, a photometric
anomaly is observed close to orbital phase $\varphi=0.50$, with an
amplitude of ${0.4^m}$ in $B$ and $V$ and ${0.3^m}$ in $R$. This
anomaly is less pronounced, or ``smeared''{} in $W$.

 This photometric peculiarity was not observed in other seasons for which we
have data, and this anomaly in the secondary minimum is also absent from other
published data. The total amplitude of the brightness variations in the 1987
secondary minimum exceeded ${0.3^m}$ in $V$, with the $V$ brightness level
close to Min~II reaching minimum values close to $13.5^m$. Such brightness
levels were not found in subsequent observations.

 Of special interest are observations of the system in the primary minimum, close
to orbital phases from $\varphi=0.97$ to $\varphi=0.03$ (we consider here a
Roche-lobe configuration with the libration point L$_2 $ located at the outer
edge of the neutron-star accretion disk). The gas flows from the system
through  L$_2 $.

 Analysis of these photometric peculiarities gives a detailed light
curve of the system in the vicinity of Min~I. In particular, we
observed a ``sharp'' minimum in the orbital light curve twice in
1987. This minimum is most likely due to the eclipse of a gaseous
structure heated by the X-ray flux at L${_2}$ [34, 43--46].

 Taking into account model constraints on the mass of the variable HZ~Her and
the neutron star, the radius of the binary orbit, the distance of the L${_2}$
from the neutron star, and the area of condensations, and assuming that the
surface brightness of gas condensations heated by X-rays close to $L_2$ can in
a first approximation be taken to be the surface brightness of HZ~Her, it is
possible that eclipses of gas condensations are observed in the primary minimum
(Min~I).

 This occurs because,  at a certain degree of Roche lobe filling by the optical
star, some gas leaves the system via L${_2}$, which is located at the outer
edge of the neutron-star accretion disk. Therefore, gas condensations can form,
probably close to L${_2}$. A blob could also form at the outer edge of the
accretion disk, with physical characteristics and optical manifestations
more or less identical to those of such a gas condensation.

 If the spatial configuration of the optical component has triangular libration
points  L${_4}$ and L${_5}$,  X-ray heated gas condensations could be located
at orbital phases $\varphi=0.166$ and $\varphi=0.833$, respectively.

 A photometric peculiarity was found in the 1987 light curve close to orbital
phases $\varphi=0.847-0.853$; this peculiarity was similar to a blob~[27, 46]
that de-excitates over some time.

\begin{figure*}[p!]
\includegraphics{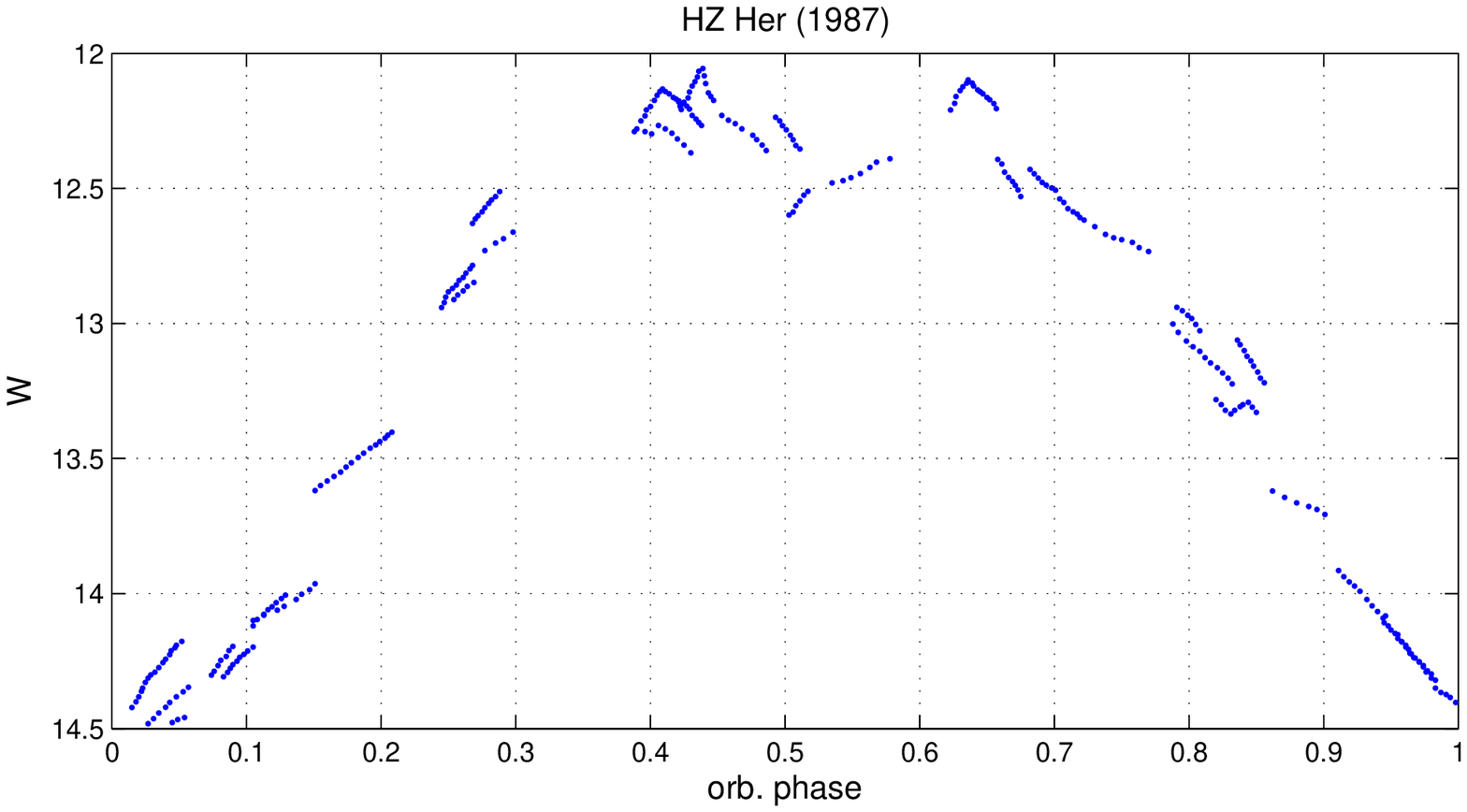}
\caption{Light curve of  HZ~Her in 1986 in  $W$
\hfill}
\end{figure*}

\begin{figure*}[p!]
\includegraphics{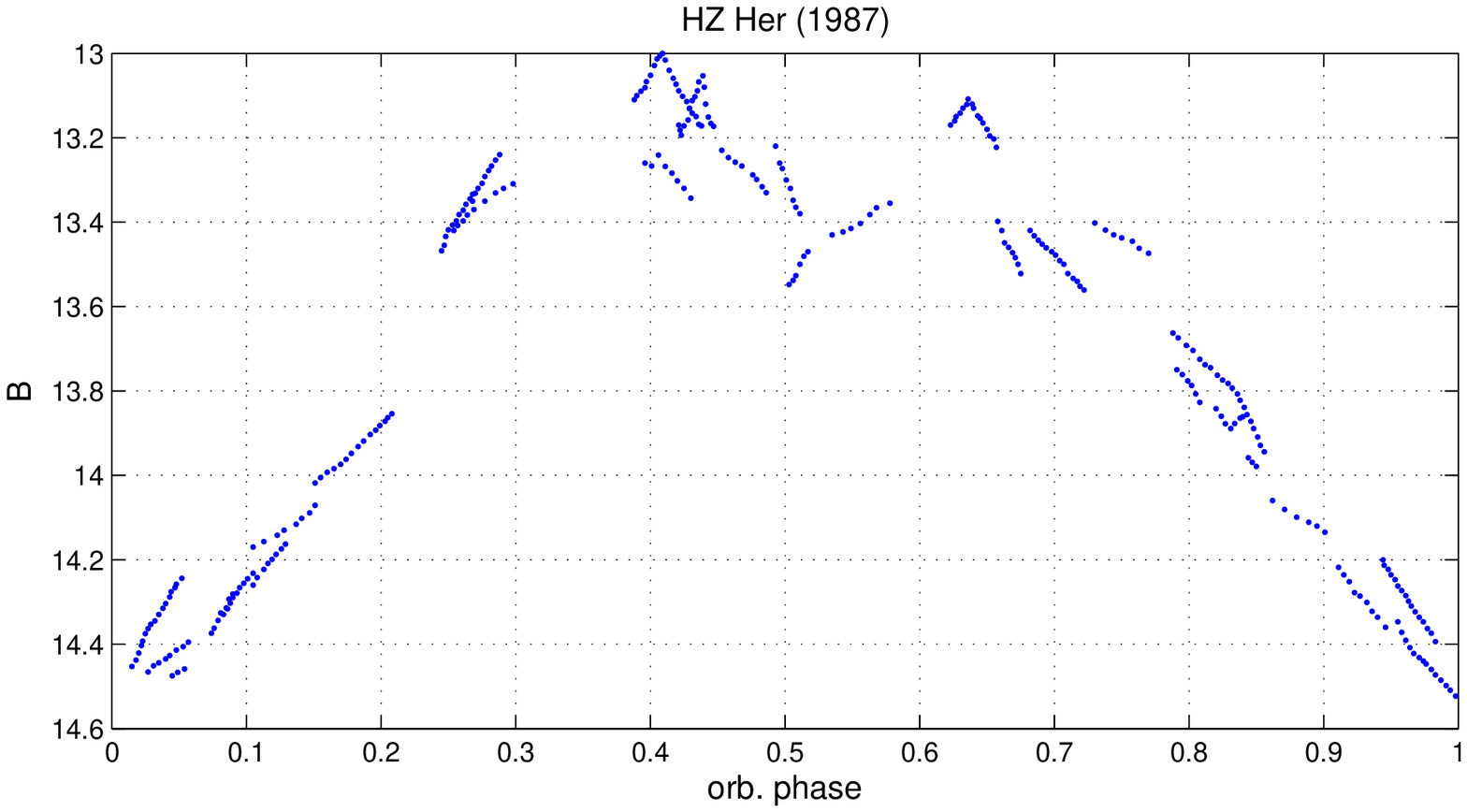}
\caption{Light curve of  HZ~Her in 1986 in  $B$
\hfill}
\end{figure*}

\begin{figure*}[p!]
\includegraphics{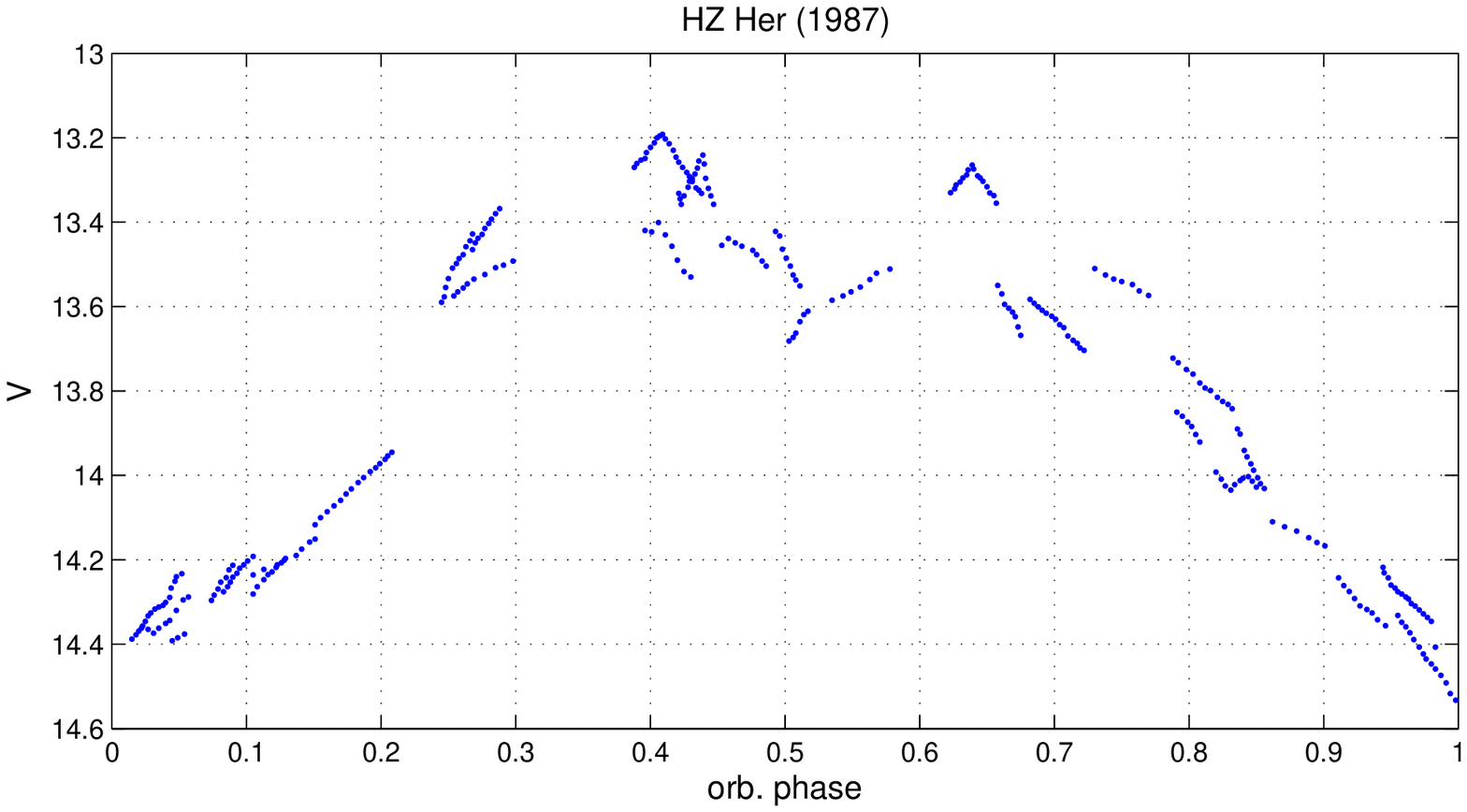}
\caption{Light curve of  HZ~Her in 1986 in  $V$
\hfill}
\end{figure*}

\begin{figure*}[p!]
\includegraphics{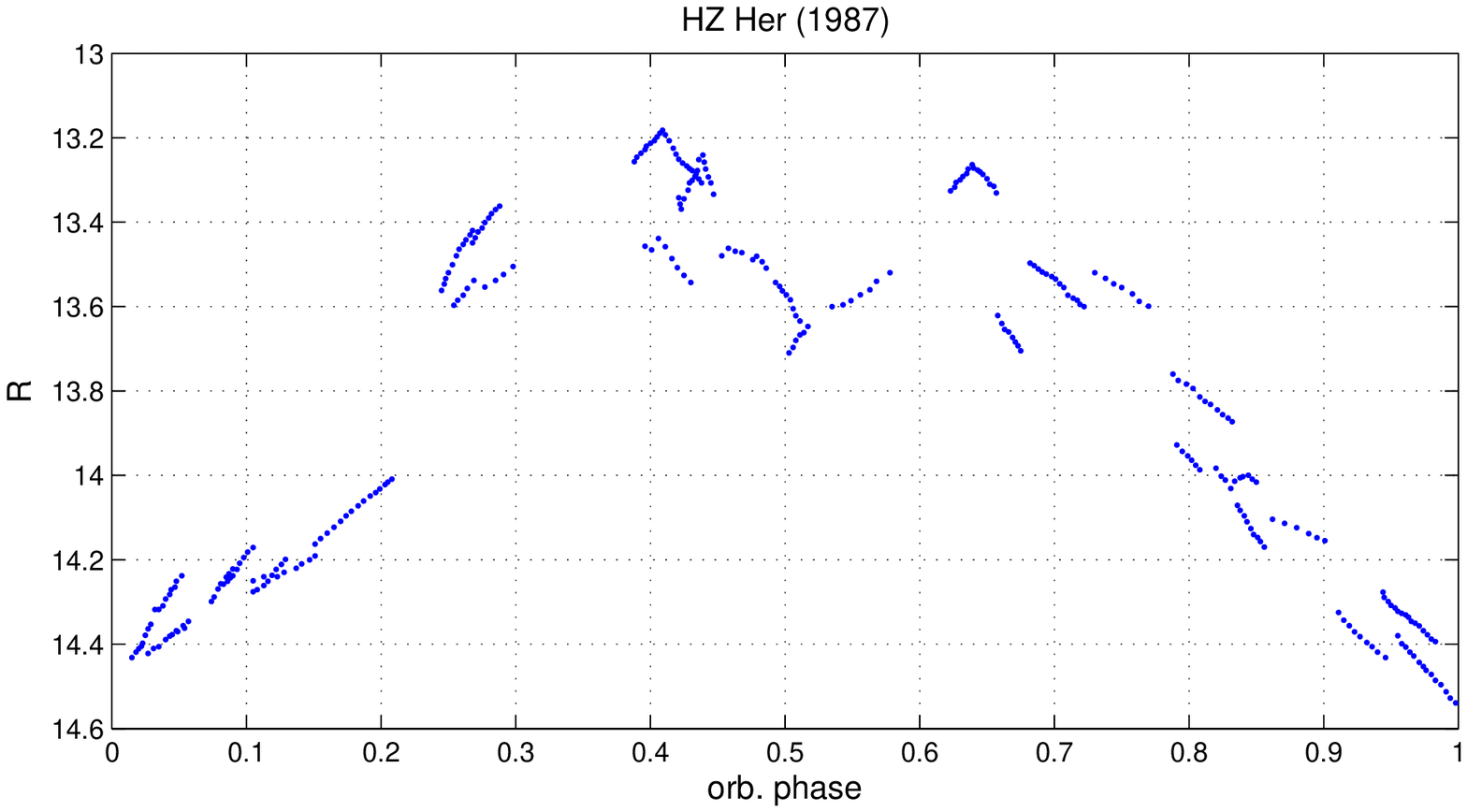}
\caption{Light curve of  HZ~Her in 1986 in $R$
\hfill}
\end{figure*}

 Deep flickering of the X-ray source on time scales of several minutes is also
observed at orbital phases close to $\varphi=-0.15$, at the epochs of dips
in the X-ray light curve~[47]. This flickering could be due to the de-excitation
of some structures, such as gas condensations in the system.

 The detailed light curve close to Min~I and Min~II shows oscillations of the
minimum brightness, $W{-}B$, and $B{-}V$ with $\varphi$, at the
level of $0.1^m$. Such oscillations were not observed in 1986 or
1987. The variations of the brightness and of the $W{-}B$,
$B{-}V$, $V{-}R$, and $B{-}R$ color indices with $\varphi$, as
well as the nature of the $(W{-}B){-}(B{-}V)$, $(V{-}R){-}(B{-}V)$
diagrams (Figs.~(6.1-6.4) and (7.1-7.2)), can be explained by
variations of the areas of accretion structures re-emitting X-ray
flux in the optical and the regimes for the gas flow from the
optical component to the neutron star.

 During X-ray eclipses, $W{-}B$  and the $W$ brightness correlate with the
accretion rate; i.e., the $W$ brightness of the system increases if there is
less gas around the neutron star. A prominent photoelectric effect in the $W$
band and variations of this effect on time scales close to ${\sim} 30{-}35$
days requires a sufficiently hot gas with temperature $T \approx2.5\times
10^6$~K and a number density of electrons $n_e = 7\times 10^{11}$~cm$^{-3}$
flowing from the inner Lagrange point $L_1$~[48]. This may form a gaseous
corona surrounding the outer regions of the accretion disk~[45].

 It is interesting to compare the light curves presented in this paper and
those obtained in other studies~[21, 32, 34, 49]. Close to elongations of the
X-ray source ($\varphi =0.25$ and $\varphi=0.75$), our observations are in
qualitative agreement with the data of other authors. This provides confirmation
that geometry of the  hot spot was constant in 1987. The highly chaotic state
of the brightness in Min~I and Min~II is striking.

 At orbital phases from $\varphi =0.30$ to $\varphi=0.40$, there is a seasonal
increase in the brightnesses in $W$ (by $0.30^m$) and $B$ (by $0.20^m$);
the $V$ and $R$ brightnesses are at their 1986 levels. There is a flat plateau
in Min~I in all the bands. Thus, in this observing season,  the left shoulder
of the light curve in the vicinity of Min~II has somewhat higher brightness
than the right shoulder. Long flares are observed in Min~I (from 20--30 to
60--80~min), with amplitudes up to $0.07^m {-} 0.08^m$. The behavior in the
$WBVR$ bands is correlated.

 Similar flares are also observed close to orbital phases 0.15--0.25 and
0.75--0.85. There are also short flares close to the phase 0.015, with
amplitudes of $0.01^m$,  $0.02^m$, $0.02^m$, and $0.03^m$ in $R$, $V$, $B$,
and $W$, respectively. These are probably related to hot condensations of matter
in the accretion structures (and in the accretion flow in general) that are
projected onto the limb of the optical component of the close binary.

\begin{figure*}[p!]
\includegraphics{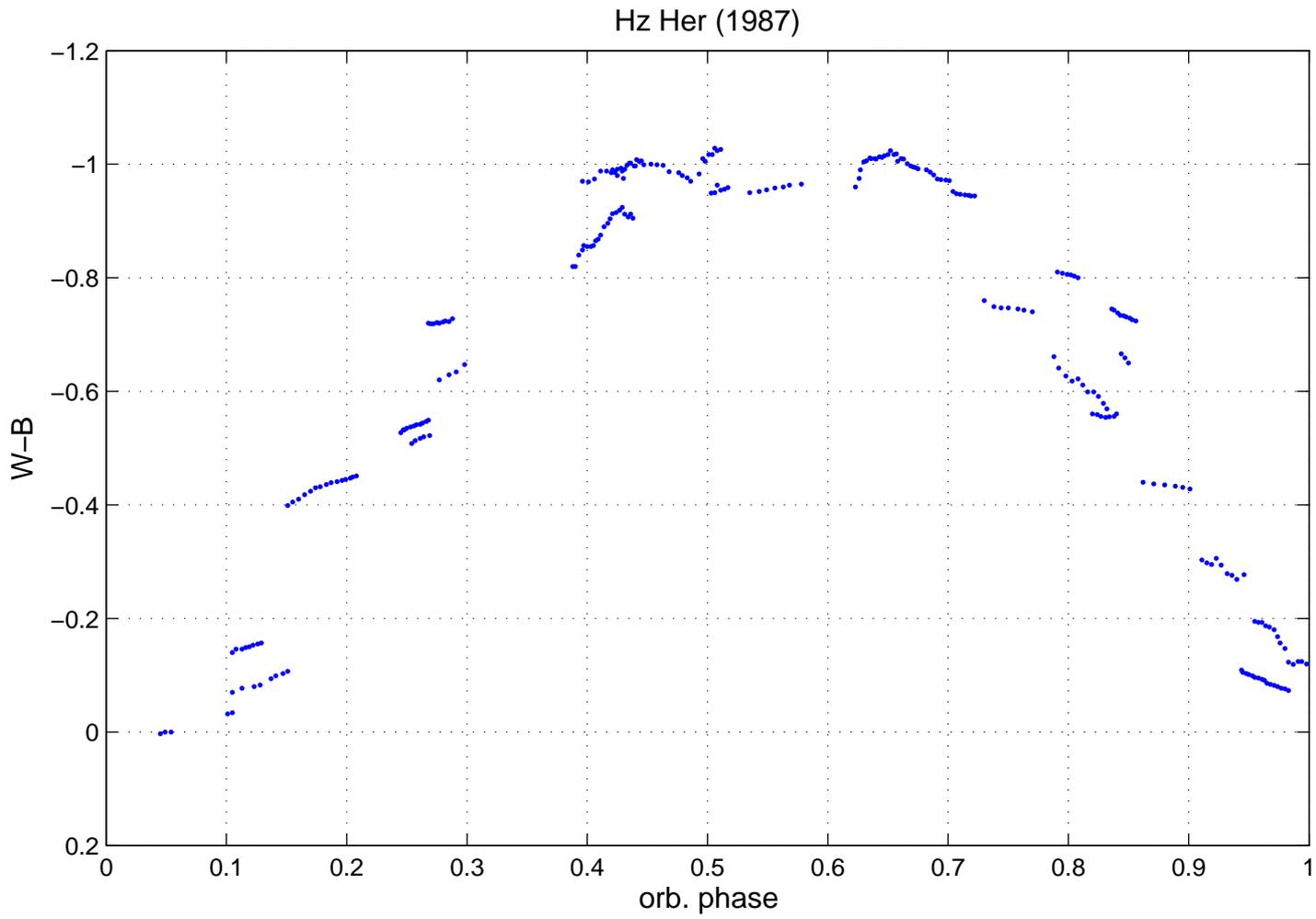}
\caption{$(W{-}B)$, vs. orbital phase $\varphi$ for the 1987
season \hfill}
\end{figure*}

\begin{figure*}[p!]
\includegraphics{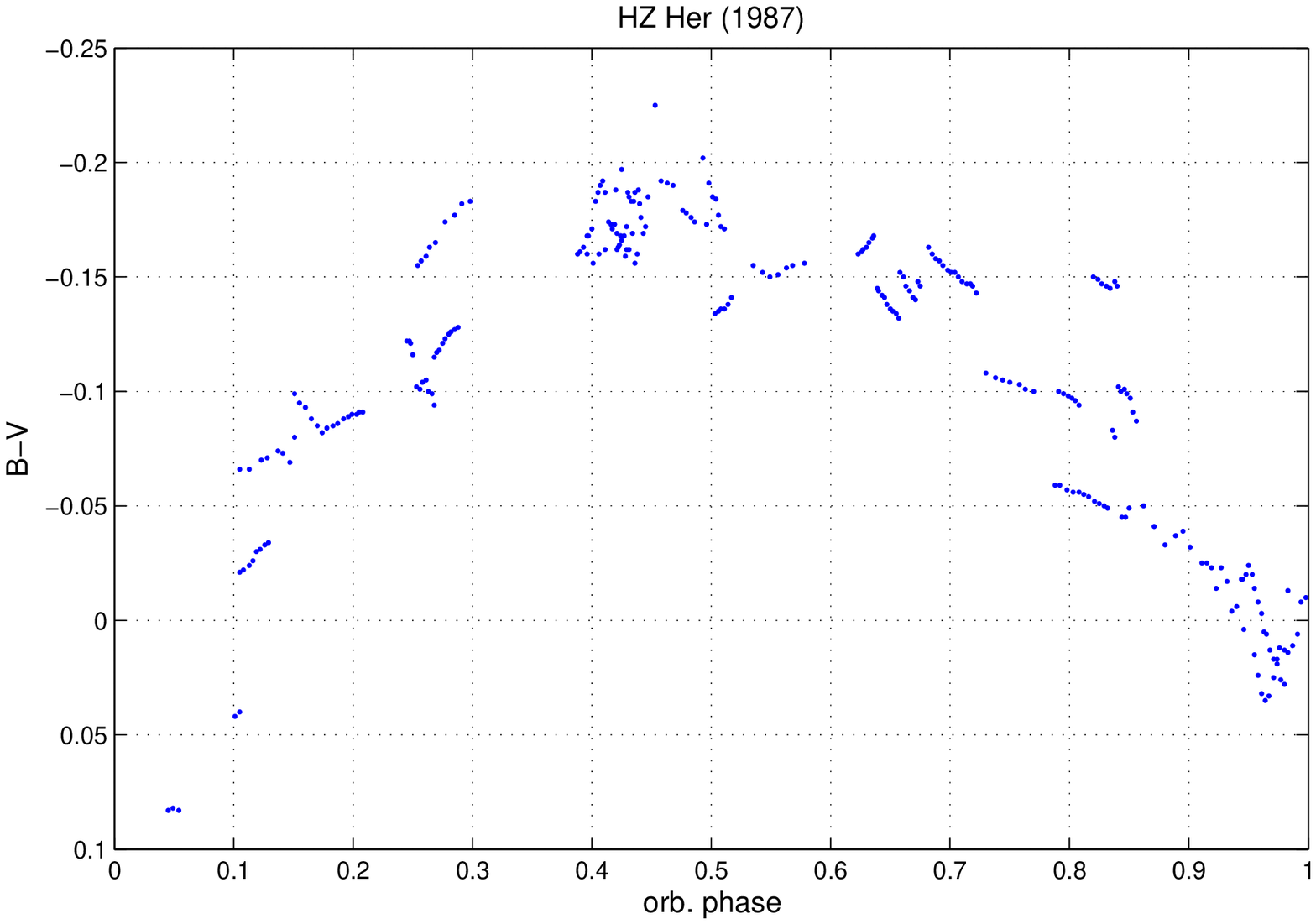}
\caption{$(B{-}V)$, vs. orbital phase $\varphi$ for the 1987
season \hfill}
\end{figure*}

\begin{figure*}[p!]
\includegraphics{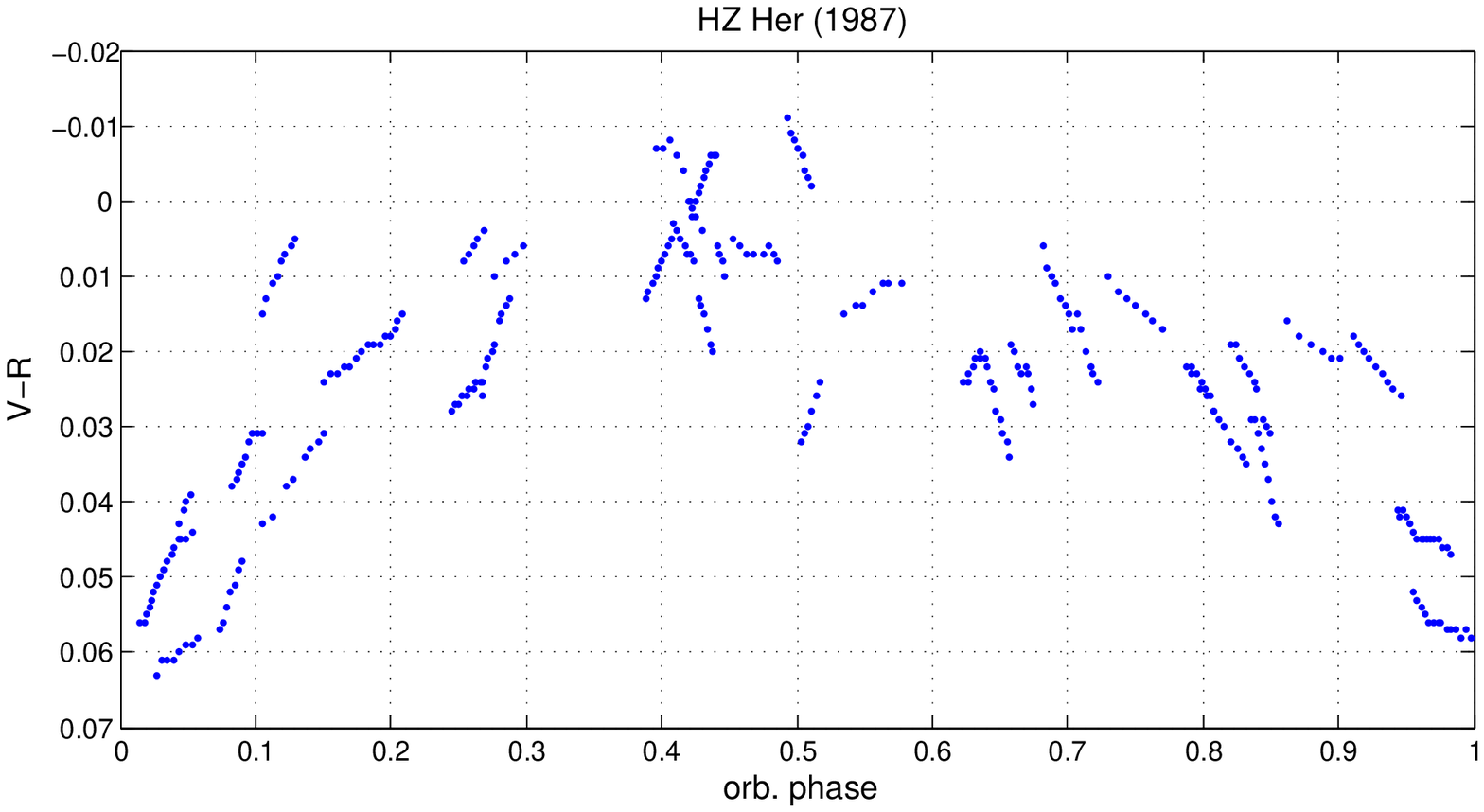}
\caption{$(V{-}R)$, vs. orbital phase $\varphi$ for the 1987
season \hfill}
\end{figure*}

\begin{figure*}[p!]
\includegraphics{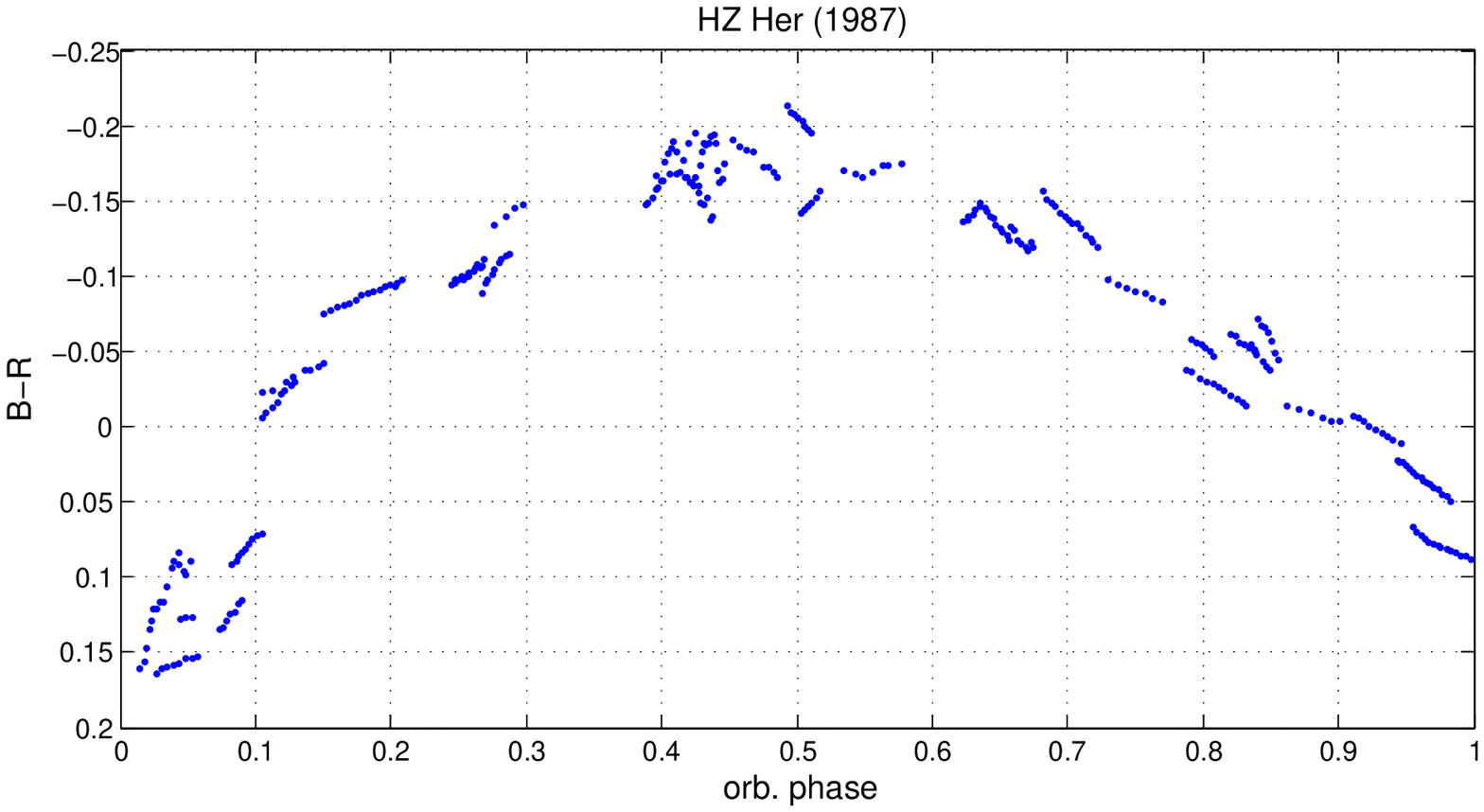}
\caption{$(B{-}R)$, vs. orbital phase $\varphi$ for the 1987
season \hfill}
\end{figure*}

\begin{figure*}[p!]
\includegraphics{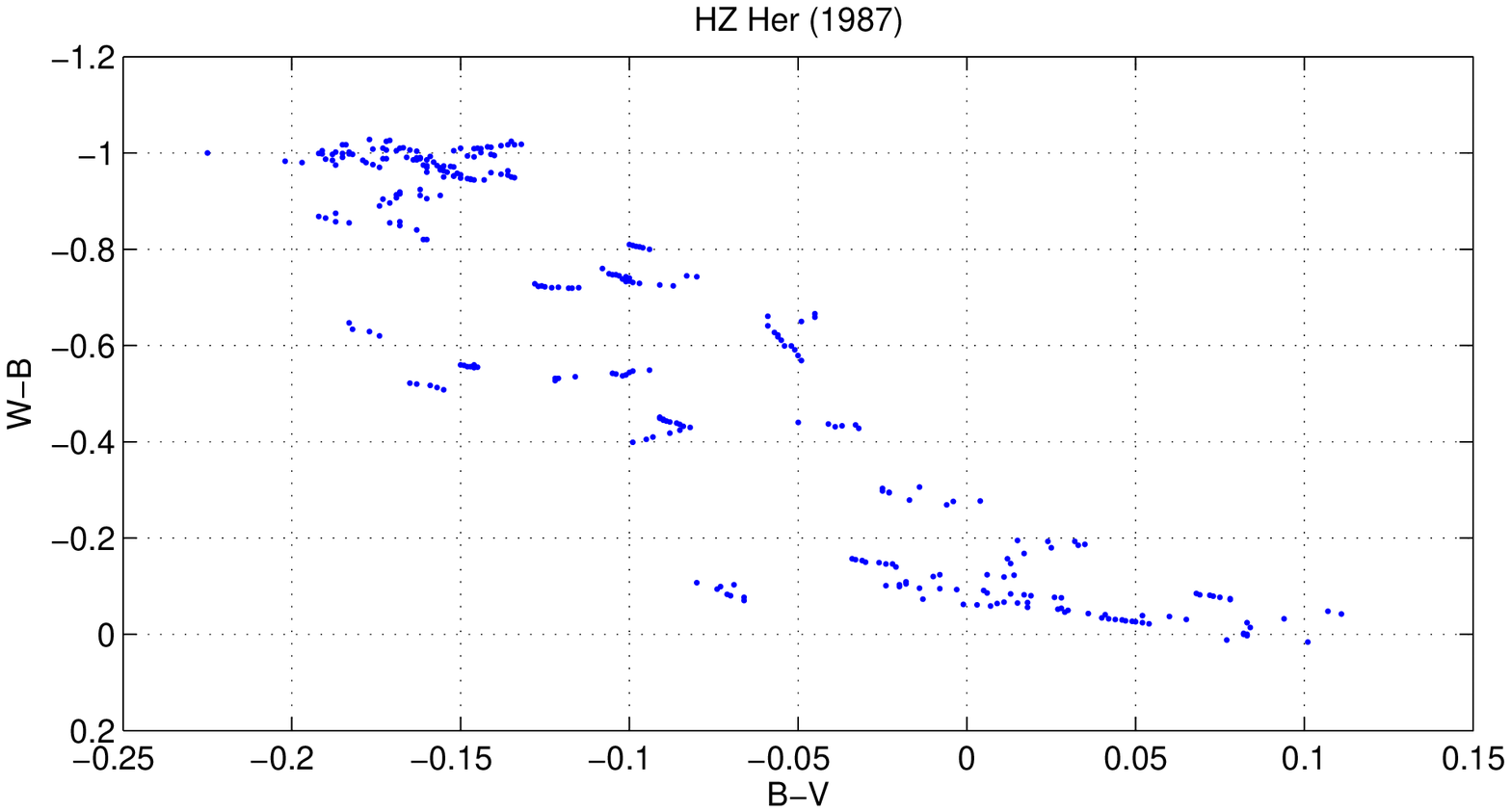}
\caption{Color--color diagrams $(W{-}B)-(B{-}V)$  for the
1987~season.
\hfill}
\end{figure*}

\begin{figure*}[p!]
\includegraphics{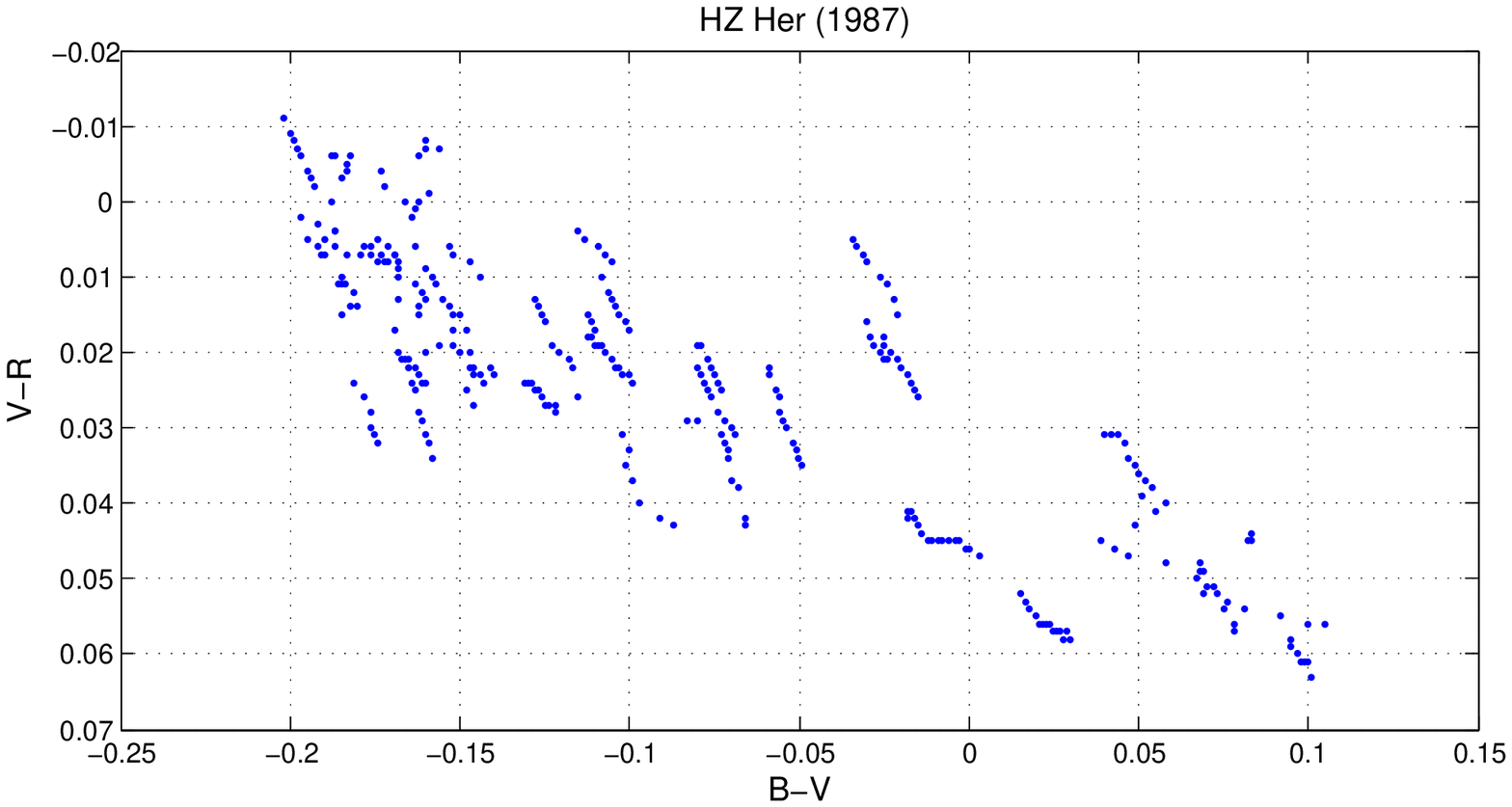}
\caption{Color--color diagrams $(V{-}R)-(B{-}V)$ for the
1986~season.
\hfill}
\end{figure*}

\section{OBSERVATIONS IN 1988}

 The results of our photoelectric observations in 1988 (Fig.~(8.1-8.4))
differ substantially from the results for 1986 and 1987, in the
presence of many components of the orbital light curve (most of
the observations were made close to the brightness maximum and in
the ``on'' {} state) and many fine photometric details in the
light curves at different phases of the 35-day precessional cycle.

 This is true first and foremost of certain peculiarities of the orbital light
curve close to the primary minimum Min~I, where a smooth (classical), flat
bottom, without any deviations, is observed in most cases. In the 1988 season,
some variations of the curve and a slight increase of the brightness were
observed close to orbital phases $\varphi = 0.97{-}0.04$.

 Such periodic variations in the light curve were also noted
in~[31, 34]. This suggests an increase in the accretion disk in
the 1988 season (the usual average size of the accretion disk is
${\approx} 0.5$ of the radius of the optical component of HZ~Her,
an A7 star). The inclination and degree of warping of the
accretion disk change~[31]. In our opinion, these variations are
related to different mass-flow regimes and occur in strictly
defined precessional phases of the 35-day cycle.

 These photometric structures in the light curve close to orbital phase
$\varphi=0.02$ form extremely infrequently. As a rule, these optical flares
are preceded by dips in the X-ray light curve, which are probably related
to variations in the rate of mass flow from the optical star HZ~Her onto
the neutron-star companion~[27, 31].

 As a consequence,  the lengths of the secondary minimum also increased. This
contradicts the conclusions drawn in~[34], and can be explained by the small
number of observational points presented in~[34] compared to our study, as
well as the absence of data in the $W$ (or $U$) and $R$ bands in~[34].
Our analysis here is based on $WBVR$ observations and a larger number of
points, especially close to Min~II. The duration of Min~II in 1988 was about
$0.41$ to $0.62$ in orbital phase, providing evidence for a periodic migration
of the hot spot along the meridian and the latitude of the optical component,
and for some increase in the accretion disk around the neutron star in 1988.

\begin{figure*}[p!]
\includegraphics{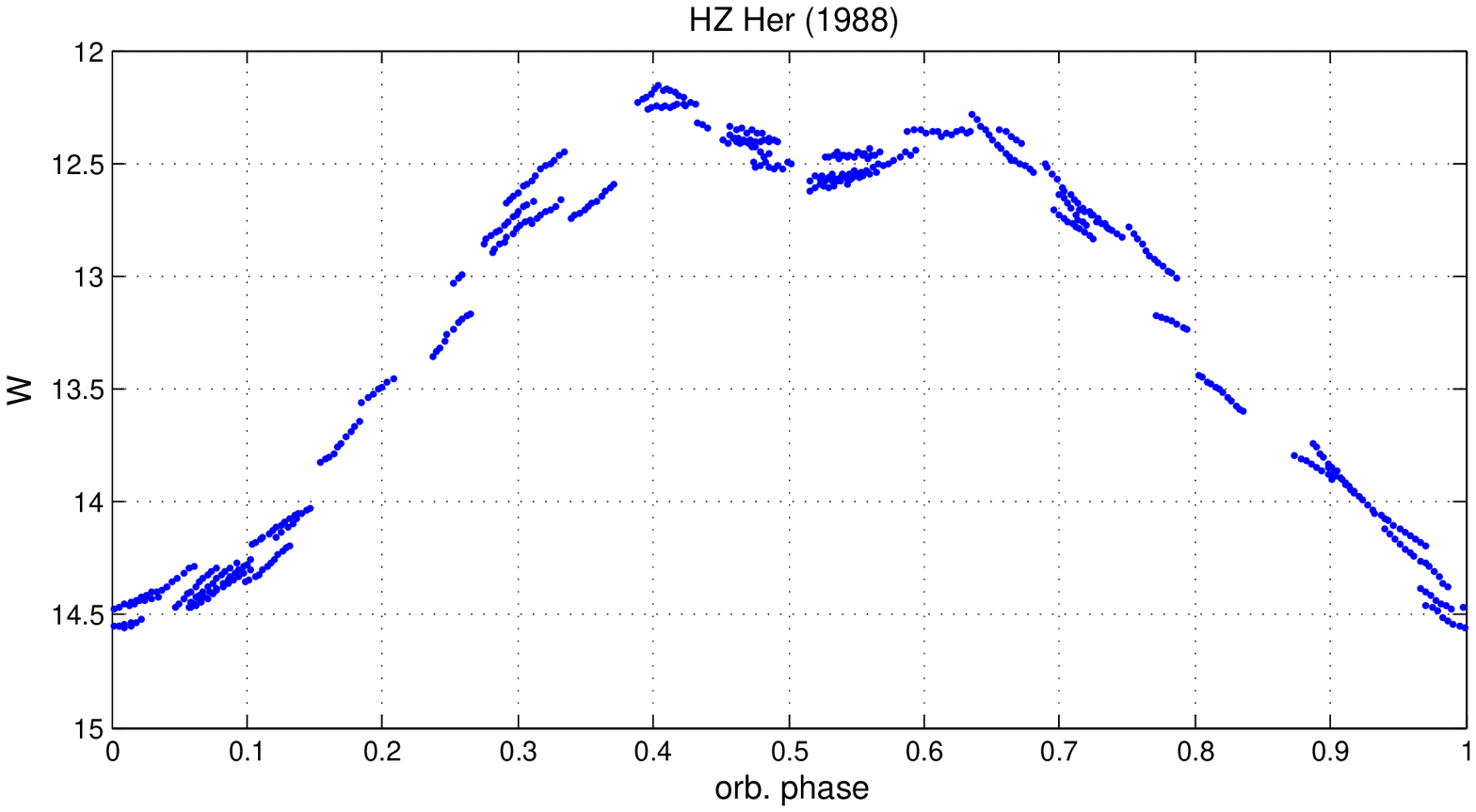}
\caption{$(W{-}B)$, vs. orbital phase $\varphi$ for the 1987
season \hfill}
\end{figure*}

\begin{figure*}[p!]
\includegraphics{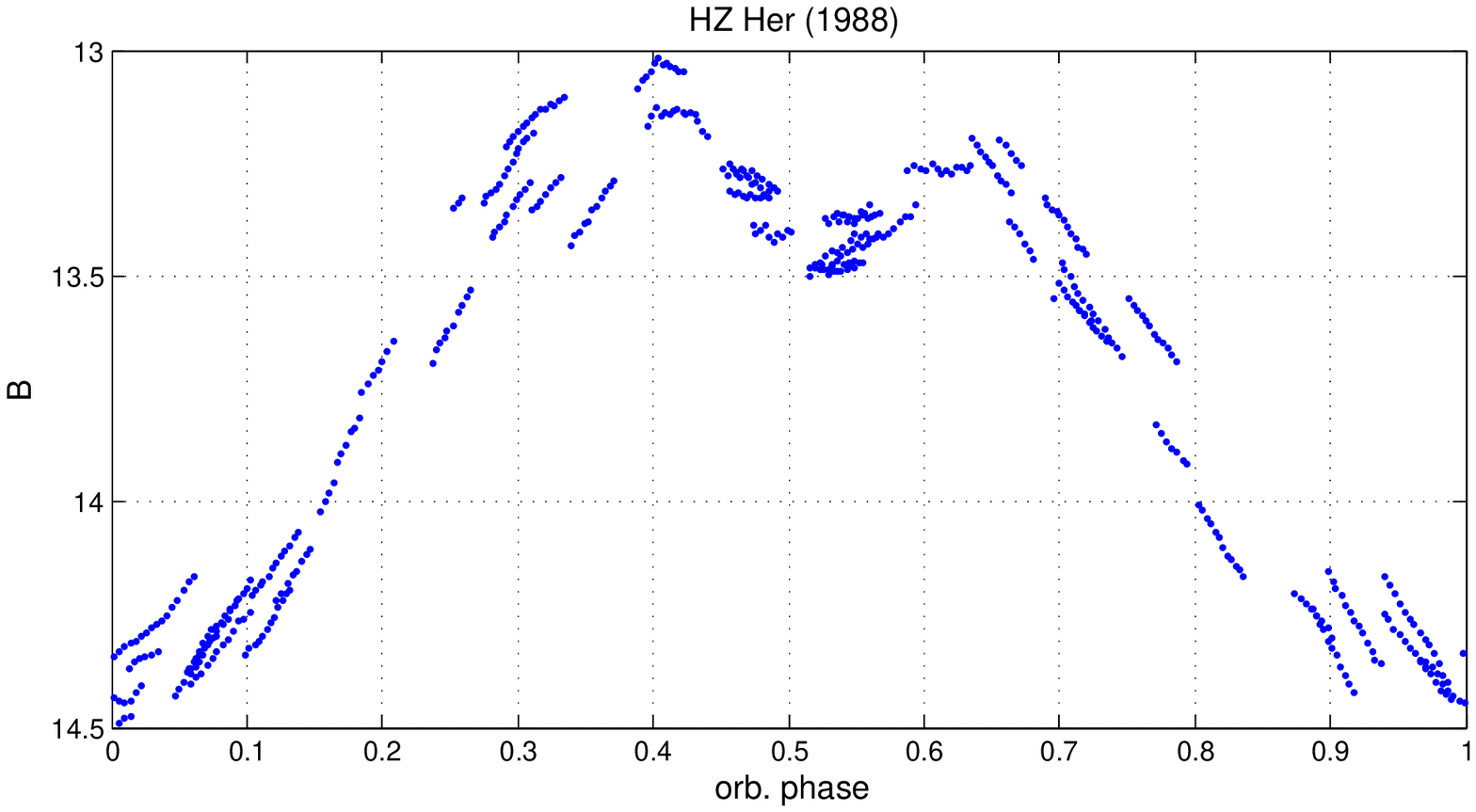}
\caption{$(W{-}B)$, vs. orbital phase $\varphi$ for the 1987
season \hfill}
\end{figure*}

\begin{figure*}[p!]
\includegraphics{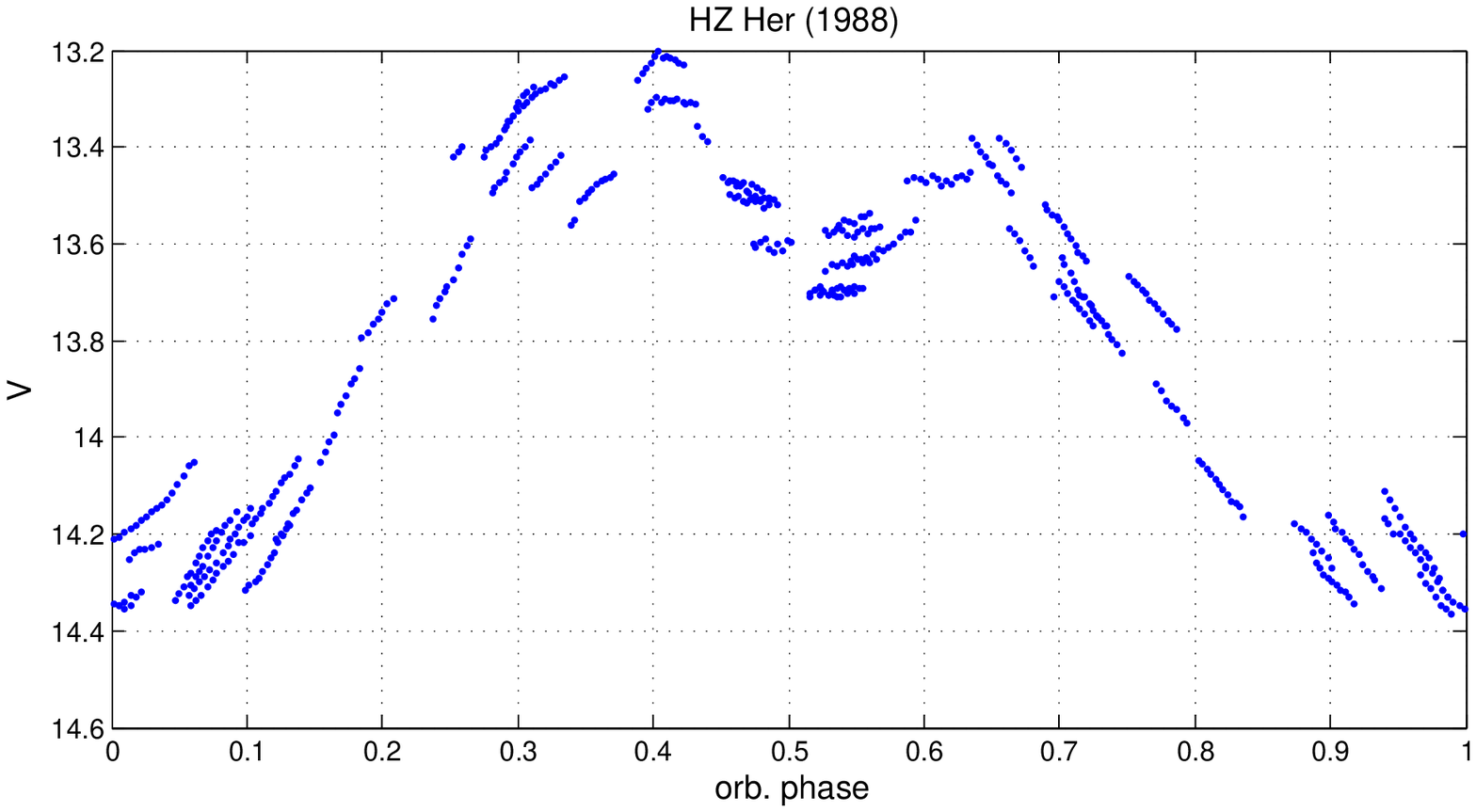}
\caption{$(W{-}B)$, vs. orbital phase $\varphi$ for the 1987
season \hfill}
\end{figure*}

\begin{figure*}[p!]
\includegraphics{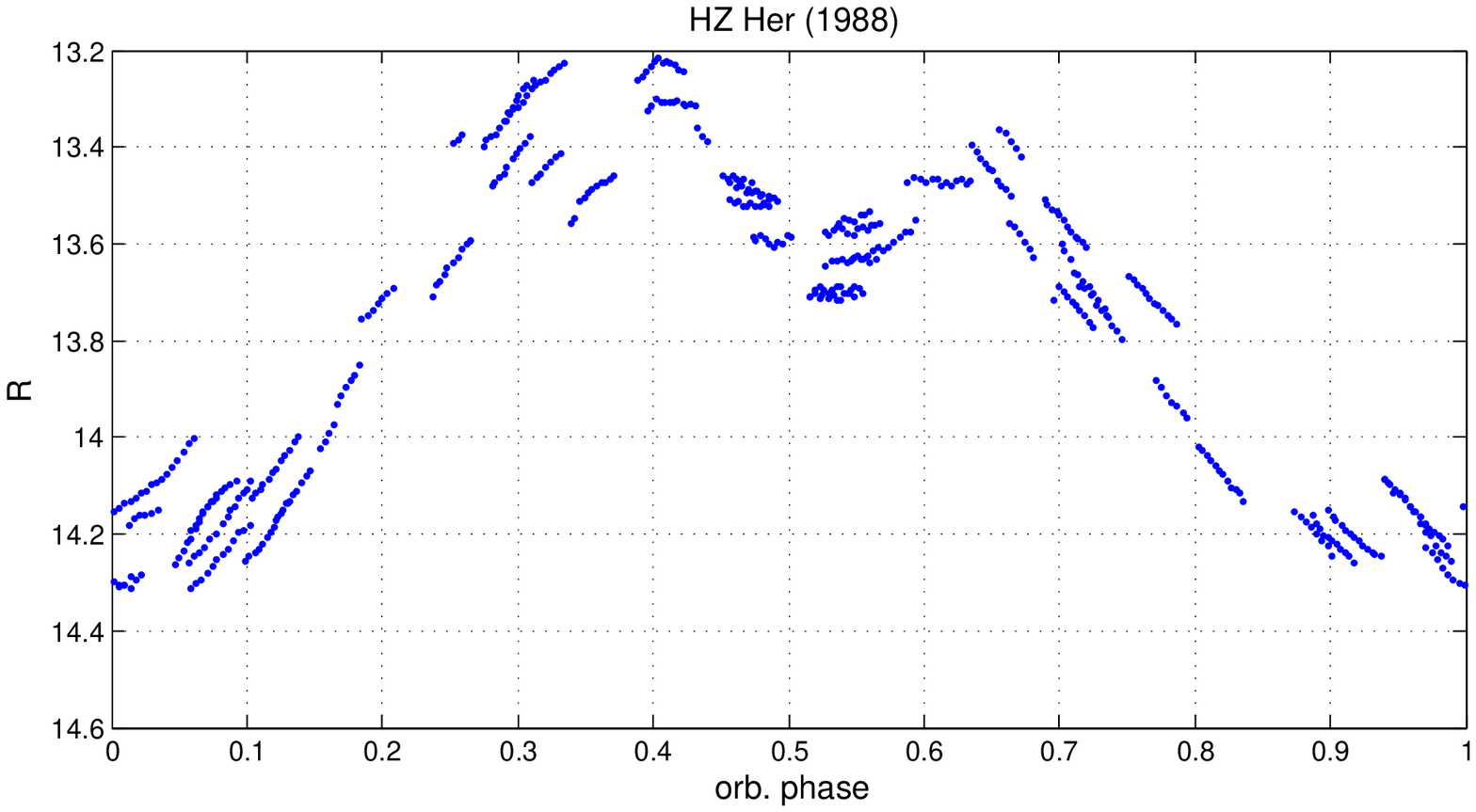}
\caption{$(W{-}B)$, vs. orbital phase $\varphi$ for the 1987
season \hfill}
\end{figure*}

\begin{figure*}[p!]
\includegraphics{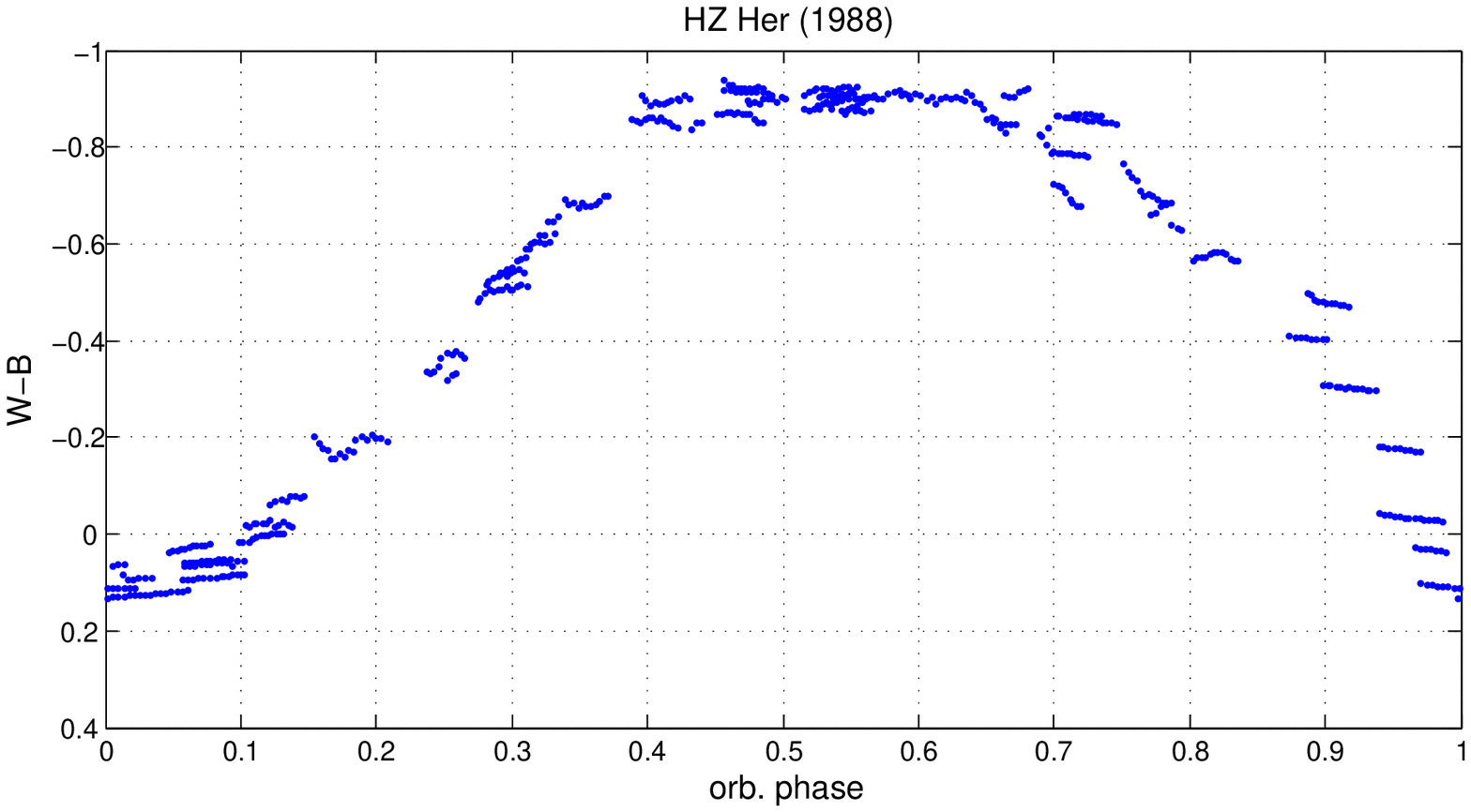}
\caption{$(W{-}B)$, vs. orbital phase $\varphi$ for the 1987
season \hfill}
\end{figure*}

\begin{figure*}[p!]
\includegraphics{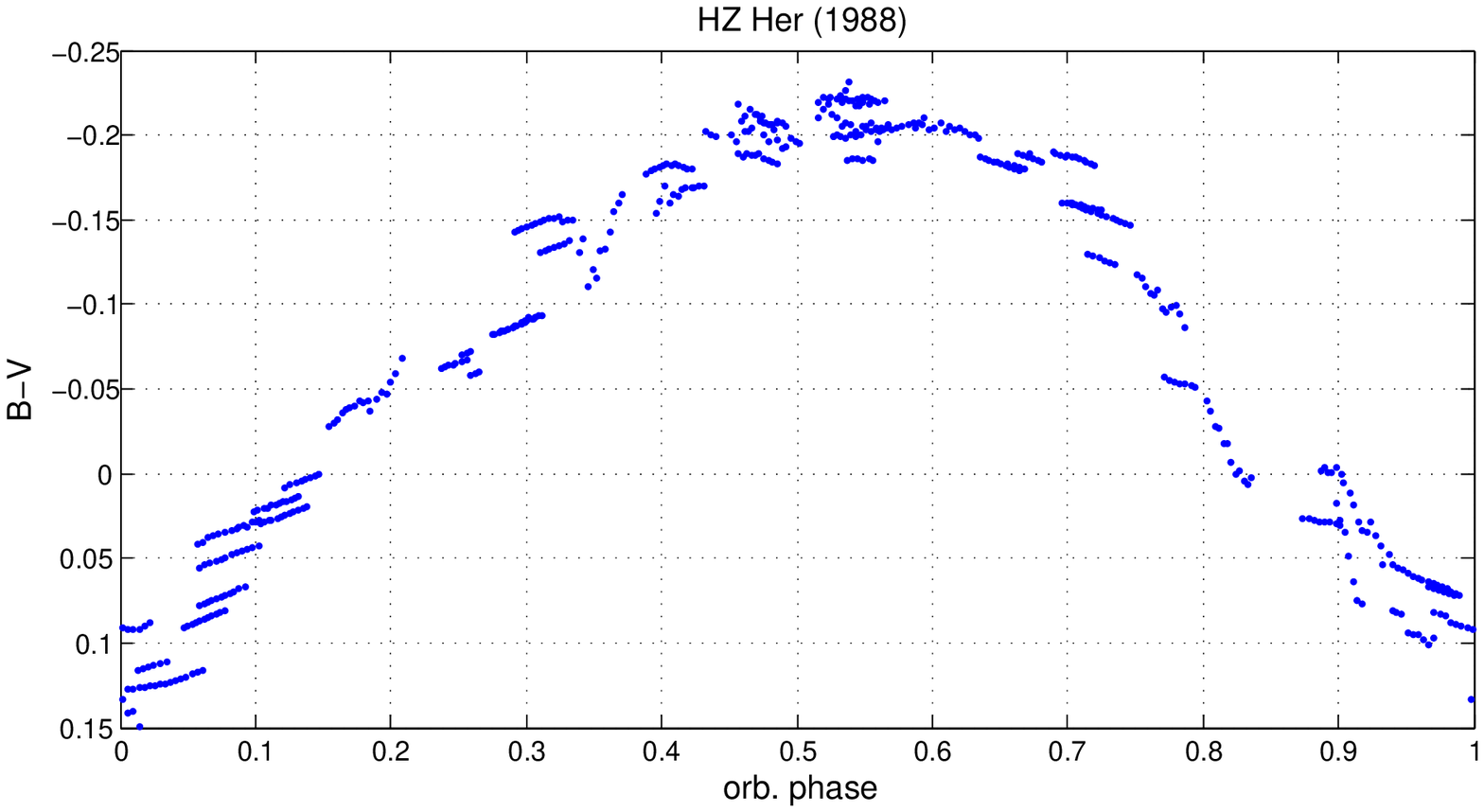}
\caption{$(W{-}B)$, vs. orbital phase $\varphi$ for the 1987
season \hfill}
\end{figure*}

\begin{figure*}[p!]
\includegraphics{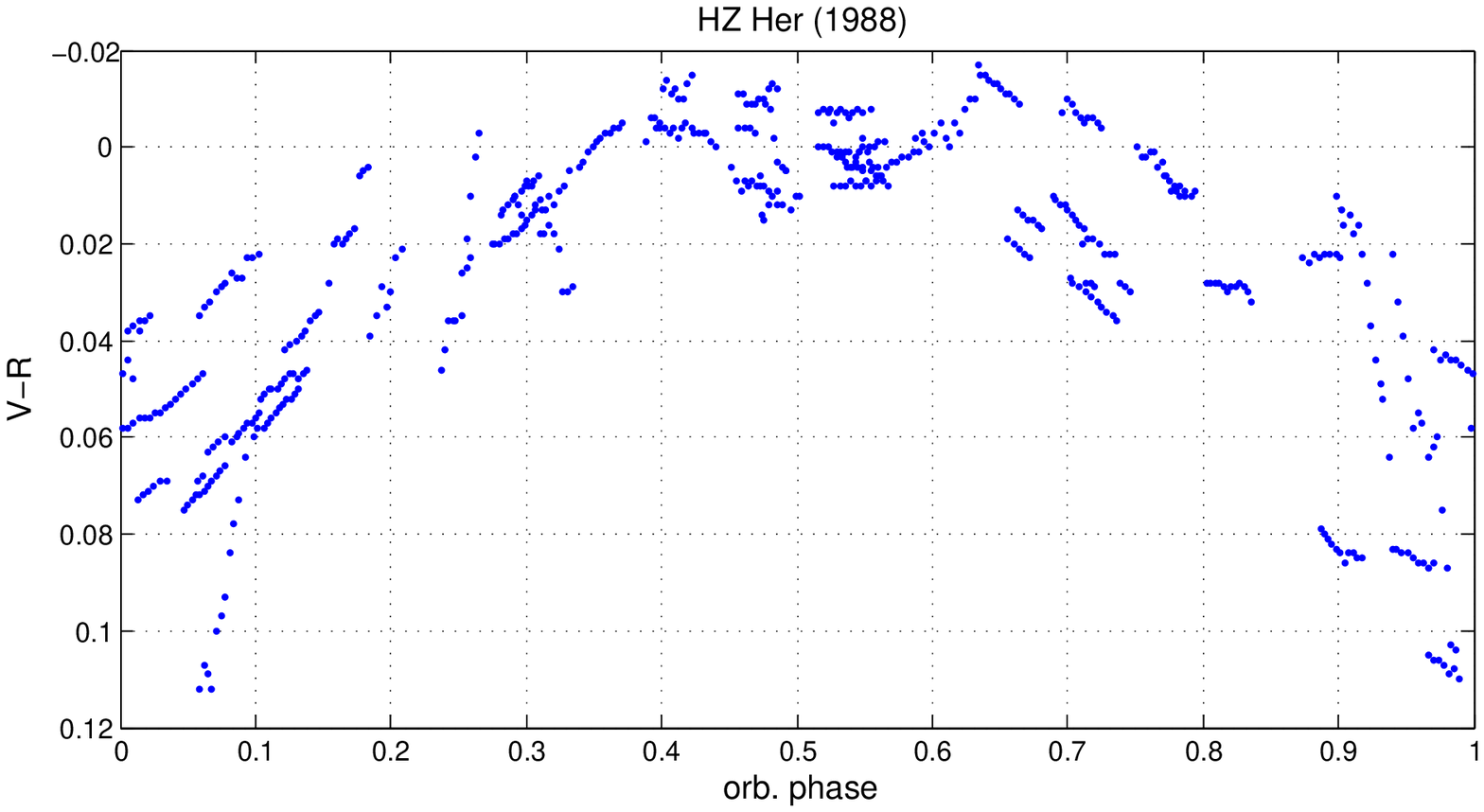}
\caption{$(W{-}B)$, vs. orbital phase $\varphi$ for the 1987
season \hfill}
\end{figure*}

\begin{figure*}[p!]
\includegraphics{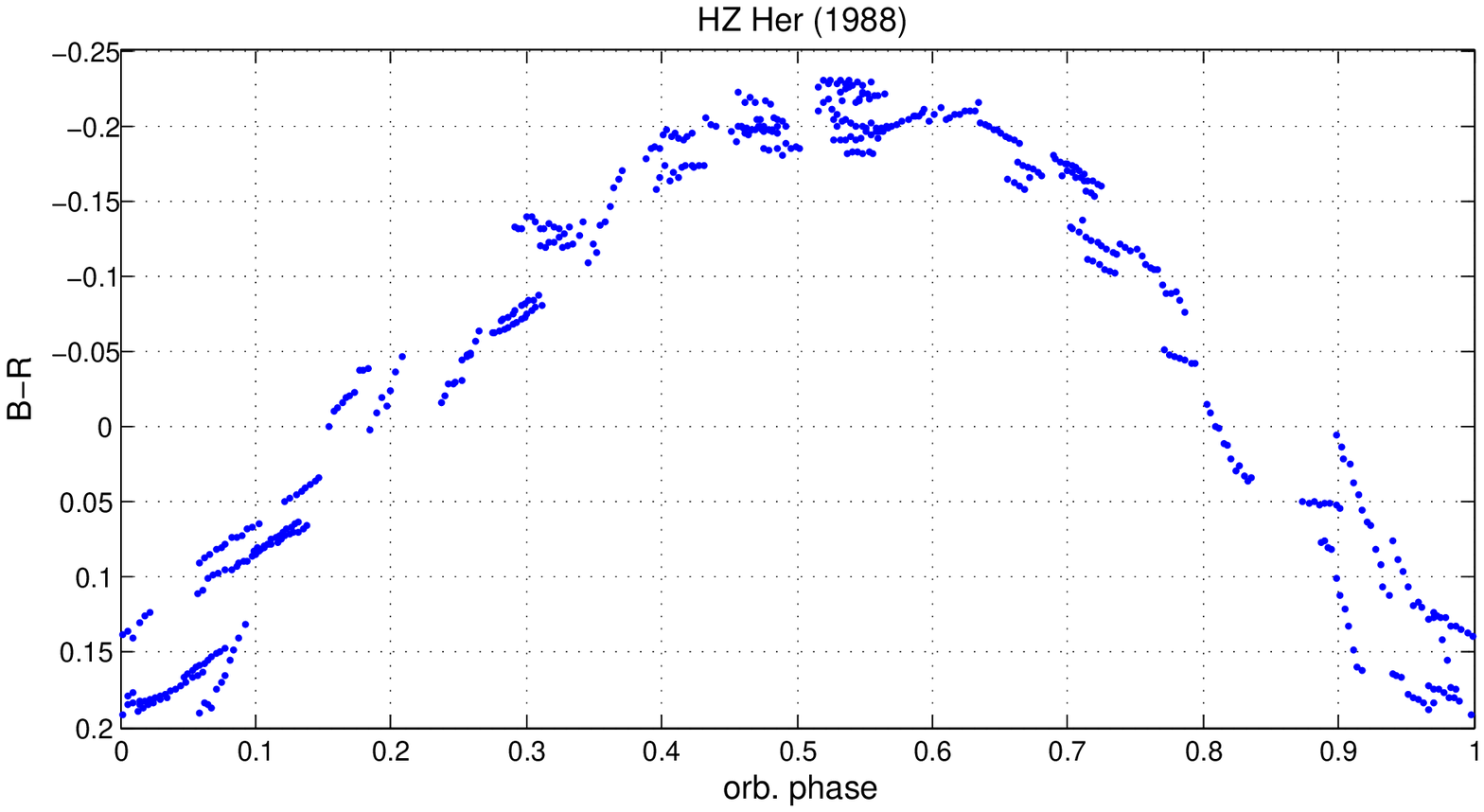}
\caption{$(W{-}B)$, vs. orbital phase $\varphi$ for the 1987
season.
\hfill}
\end{figure*}

\begin{figure*}[p!]
\includegraphics{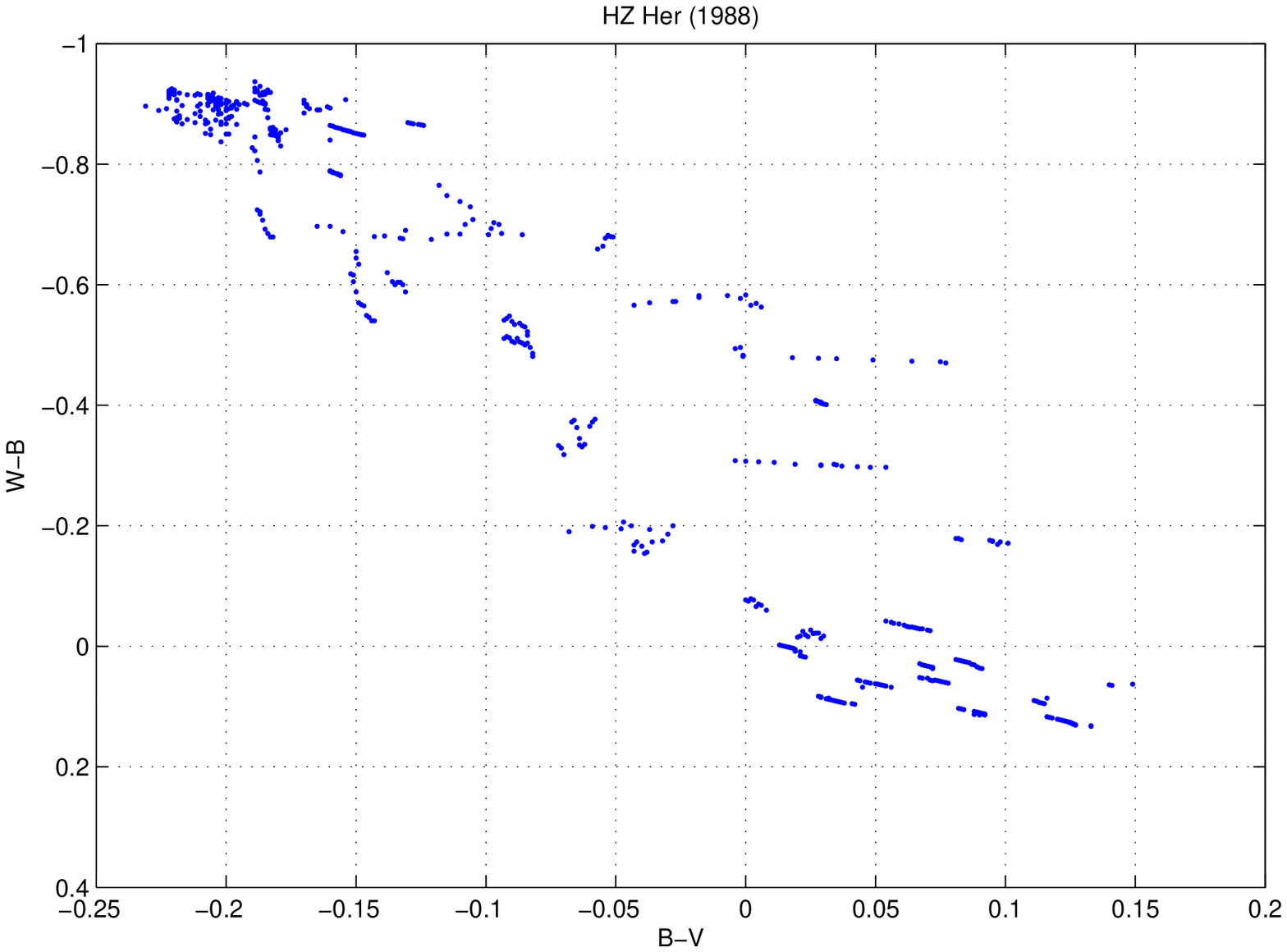}
\caption{Color--color diagrams  $(W{-}B)-(B{-}V)$ for the
1988~season.
\hfill}
\end{figure*}

\begin{figure*}[p!]
\includegraphics{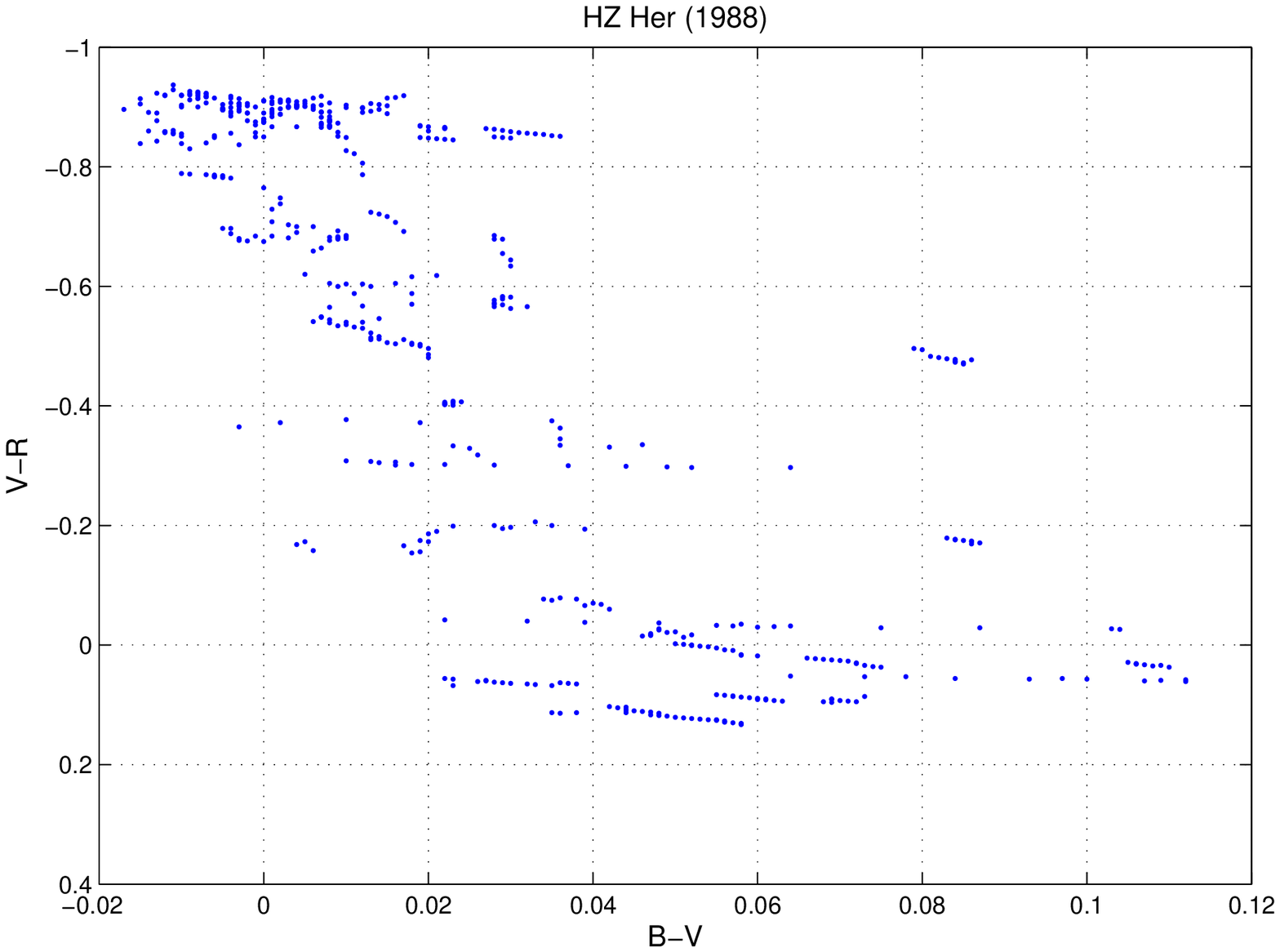}
\caption{Color--color diagrams $(V{-}R)-(B{-}V)$ for the
1988~season.
\hfill}
\end{figure*}

 Also in the 1988 season, the minimum values of $W{-}B \approx0$ were detected
(at precession phase  $\psi=0.86$; Fig.~(9.1-9.4)), which
correspond to the spectral class of the optical component in Min~I
and a deficit of UV emission. Note that all photometric effects
related to the 35-day cycle are manifest most strongly in the UV,
as is clearly visible in the plots for $W$ and $B$ for all the
observing seasons.

 A joint and qualitative analysis of our observational data, X-ray
observations~[50--52], and observations in other spectral bands (both in the
same seasons and later) enables refinement of the geometry and places limits on
variations of the hot spot and accretion structures. In the studies cited
above, the ${35^{\textrm{d}}}$ period was analyzed at precessional phases
$\psi=0.76{-}0.80$ with the aim of looking for manifestations of blobs close
to the gas flow projected onto the orbital plane of the system. A similar
analysis was carried out in~[34]. $\textrm{HZ~Her} = \textrm{Her~X-1}$ was
observed in the anomalous low state at phase $\psi=0.76{-}0.88$~[53].

 The total brightness amplitude close to Min~II in the 1988 season was up to
${\sim} 0.55^m$ in $V$. The brightness level close to Min~II (orbital phase
$0.475 {-} 0.485$) reached its minimum values: $12.6^m$ in $W$, $13.5^m$ in
$B$, $13.4^m$ in $V$, $13.4^m$ in $R$.

 $W{-}B$ was small compared to the previous observing seasons. Analysis of
the $(W{-}B) {-}(B{-}V)$, $(V{-}R) {-}(B{-}V)$ color--color
diagrams (Fig.~(10.1-10.2) suggests that the hot spot was located
along the trajectory of the motion of the accretion structures
over the limb of the optical component of the close binary, but
the relation with the trajectory was less pronounced than in 1986
and 1987.

 Thus, some chaos of the brightness close to Min~II was observed in the 1988
season at various phases of the 35-day cycle, especially in the UV. This
may indicate spatial evolution of the optically thick and warped accretion
disk around the neutron star. The disk creates shadows on the surface of the
optical component, which can give rise to physical variability of the accretion
structures and  manifestations of gas flows in the system; in turn, these
shadows come about due to differences in the rate of mass flow from the optical
component onto the neutron star.

\section{CONCLUSIONS}

Our study leads to the following conclusions.

-- We have presented data on eight full precessional cycles of the close binary,
enabling us to identify certain fine photometric effects that both confirm the
conclusions of other authors and provide new information on the system.

-- The homogeneity of our set of data containing electophotometric observations
in four optical bands ($W$, $B$, $V$, $R$) makes them of special value, and
provides additional information on fine effects in the $\textrm{HZ~Her} =
\textrm{Her~X-1}$ system.

-- In the 1986--1988 observing seasons (and subsequent 1989--1998 seasons),
there is a clear correlation between the light curves in all four bands
($W$, $B$, $V$, $R$).

-- The light curves of HZ~Her we have presented here are in qualitative
agreement with light curves obtained by other authors from photometric
observations taken in the same periods, and the data correlate well at all
phases of the precessional cycle. This correlation~[31] is also confirmed
by the location of the outer regions of the accretion disk during the ingress
and egress of the X-ray eclipse, as follows from the $(W{-}B) {-} (B{-}V)$
color--color diagrams.

-- The $W{-}B$, $B{-}V$, $V{-}R$, and $B{-}R$ color indices differ substantially
at phases of the X-ray eclipse, $0.930 {-} 0.078$. This can probably be
explained by season-to-season changes in the physical state of the neutron-star
accretion disk and the conditions for mass flow in the close binary.

\section{ACKNOWLEDGMENTS}

 The author thanks the Supervisor of the Galactic and Variable Star Group
of the Sternberg Astronomical Institute A.S.~Rastorguev for providing the
possibility of carrying out this study and for  scientific advice on various
questions pertaining to the presentation of the paper. The author also thanks
N.I.~Shakura, A.I.~Zakharov, and E.K.~Scheffer for useful discussions and
valuable comments.

\end{document}